\documentclass[format=acmlarge, review=false, screen=true]{acmart}

\usepackage{booktabs} 

\acmJournal{TKDD} 
\acmVolume{0}
\acmNumber{1}
\acmArticle{00}
\acmYear{2018}
\acmMonth{8}

\setcopyright{acmcopyright}

\acmDOI{0000001.0000001}

\usepackage{multirow}
\usepackage[noend]{algpseudocode}
\usepackage[ruled]{algorithm2e} 

\SetAlFnt{\small}
\SetAlCapFnt{\small}
\SetAlCapNameFnt{\small}
\SetAlCapHSkip{0pt}
\IncMargin{-\parindent}

\usepackage{array}
\newcolumntype{P}[1]{>{\centering\arraybackslash}p{#1}}
\newcolumntype{M}[1]{>{\centering\arraybackslash}m{#1}}
\newcolumntype{R}[1]{>{\arraybackslash}m{#1}}

\definecolor{orange}{rgb}{1,0.5,0}
\definecolor{graynode}{RGB}{20,20,20}
\definecolor{crimsonred}{RGB}{220,20,60}
\definecolor{darkgraynode}{gray}{0.5}
\definecolor{lightgraynode}{gray}{0.8}

\usepackage{rotate}
\usepackage{capt-of}
\usepackage{tabulary}
\usepackage{setspace}
\usepackage{amssymb}
\usepackage{pifont}

\definecolor{gray}{RGB}{20,20,20}
\definecolor{gray}{RGB}{0.7,0.7,0.7}
\definecolor{greencm}{RGB}{0,153,0}

\definecolor{plotblue}{RGB}	{30,144,255}
\definecolor{plotgreen}{RGB}	{50,205,50}
\definecolor{plotred}{RGB}	{220,20,60}

\definecolor{myyellow}{RGB}{255,255,204}
\definecolor{myred}{RGB}{255,204,204}
\definecolor{myblue}{RGB}{0,200,255}
\definecolor{mygreen}{RGB}{80,220,80}

\newcommand*\hrulefillvar[1][0.4pt]{\leavevmode\leaders\hrule height#1\hfill\kern0pt}

\usepackage{ragged2e}

\usepackage{comment}
\usepackage{tabularx}

\usepackage{subfigure}
\usepackage{enumitem}

\definecolor{thedarkblue}{RGB}{0,0,120} 
\definecolor{mydarkblue}{rgb}{0,0.08,0.45} 

\usepackage{hyperref}
\hypersetup{%
    colorlinks=true,
    linkcolor=mydarkblue,
    citecolor=mydarkblue,
    filecolor=mydarkblue,
    urlcolor=mydarkblue}
    
\usepackage{microtype}
\usepackage{graphicx}

\usepackage{balance}
\usepackage{tabularx}

\usepackage{booktabs}
\usepackage{tabularx}
\usepackage{comment}

\begin{document}

\title{A Survey of Parallel Sequential Pattern Mining}

\author{Wensheng Gan}
\affiliation{%
	\institution{Harbin Institute of Technology (Shenzhen)}
	\city{Shenzhen}
	\country{China}
}
\email{wsgan001@gmail.com}

\author{Jerry Chun-Wei Lin}
\authornote{This is the corresponding author}
\affiliation{%
	\institution{Harbin Institute of Technology (Shenzhen)}
	\city{Shenzhen}
	\country{China}
}
\email{jerrylin@ieee.org}

\author{Philippe Fournier-Viger}
\affiliation{%
	\institution{Harbin Institute of Technology (Shenzhen)}
	\city{Shenzhen}
	\country{China}	
}
\email{philfv@hitsz.edu.cn}

\author{Han-Chieh Chao}
\affiliation{%
	\institution{National Dong Hwa University}
	\city{Hualien}
	\country{Taiwan}
}
\email{hcc@ndhu.edu.tw}

\author{Philip S. Yu}
\affiliation{%
	\institution{University of Illinons at Chicago}
	\city{Chicago}
	\country{USA}
}
\email{psyu@uic.edu}

\renewcommand\shortauthors{W. Gan et al.}

\begin{abstract}

With the growing popularity of shared resources, large volumes of complex data of different types are collected automatically. Traditional data mining algorithms generally have problems and challenges including huge memory cost, low processing speed, and inadequate hard disk space. As a fundamental task of data mining, sequential pattern mining (SPM) is used in a wide variety of real-life applications. However, it is more complex and challenging than other pattern mining tasks, i.e., frequent itemset mining and association rule mining, and also suffers from the above challenges when handling the large-scale data. To solve these problems, mining sequential patterns in a parallel or distributed computing environment has emerged as an important issue with many applications. In this paper, an in-depth survey of the current status of parallel sequential pattern mining (PSPM) is investigated and provided, including detailed categorization of traditional serial SPM approaches, and state of the art parallel SPM. We review the related work of parallel sequential pattern mining in detail, including partition-based algorithms for PSPM, Apriori-based PSPM, pattern growth based PSPM, and hybrid algorithms for PSPM, and provide deep description (i.e., characteristics, advantages, disadvantages and summarization) of these parallel approaches of PSPM. Some advanced topics for PSPM, including parallel quantitative / weighted / utility sequential pattern mining, PSPM from uncertain data and stream data, hardware acceleration for PSPM, are further reviewed in details. Besides, we review and provide some well-known open-source software of PSPM. Finally, we summarize some challenges and opportunities of PSPM in the big data era.

\end{abstract}



\maketitle

\section{Introduction}
\label{sec:introduction}

With the rapid development of information technology and data collection, Knowledge Discovery in Databases (KDD), which is also called data mining provides a powerful capability to discover meaningful and useful information from different types of complex data \cite{agrawal1994fast,agrawal1995mining, agrawal1993mining, han2004mining, srikant1996mining}. KDD has numerous real-life applications and is crucial to some of the most fundamental tasks such as  frequent itemset mining \cite{han2004mining, fournier2017survey0}, association rule mining \cite{agrawal1993mining,zaki2000scalable}, sequential pattern mining \cite{pei2004mining, srikant1996mining, fournier2017survey,van2018mining,yun2008new}, clustering \cite{berkhin2006survey, jarvis1973clustering}, classification \cite{quinlan2014c4,nguyen2012classification}, and outline detection \cite{lee2000adaptive}.

Most traditional data mining algorithms are designed to run on a single computer (node), and are thus called single-node techniques. They can discover various kinds of patterns from various types of databases. At the same time, in recent decades, data mining has been studied extensively and applied widely \cite{agrawal1994fast, agrawal1993mining, berkhin2006survey, chen1996data, geng2006interestingness, han2004mining, yan2002gspan}. These techniques perform well on small datasets, however, due to the limited memory capacity and computation capability of a single node, these data mining methods become inefficient over big data. The memory requirements for handling the complete set of desired results increase quickly, and the computational cost can be expensive on a single machine. All aforementioned methods are serialized. When handling large-scale data, these methods are fundamentally inappropriate due to many reasons, including the huge amounts of data, infeasibility of bandwidth limitation, as well as the fact that larger inputs demands parallel processing, and privacy concerns. Unfortunately, parallelization of the mining process is a difficult task. It is an important issue to develop more adaptable and flexible mining frameworks.

Parallel data mining (PDM) \cite{zaki1999parallel,gan2017data} is a type of computing architecture in which several processors execute or process an application. Research and development work in the area of parallel data mining concerns the study and definition of parallel mining architectures, methods, and tools for the extraction of novel, useful, and implicit patterns from databases using a high-performance architecture. In some cases, PDM distributes the mining computation w.r.t. multi-core over more than one node. When data mining tools are implemented on high-performance parallel computers, they can analyze massive databases in parallel processing within a reasonable time. In general, parallel computation allows for solving larger problems and executing applications that are parallel and distributed in nature. Parallel computing is becoming increasingly common and used to accelerate processing of the massive amount of data. So far, some parallelized Apriori-like algorithms have been implemented with the MapReduce framework \cite{dean2010mapreduce} and achieved certain speedup compared with single-node methods. However, some previous studies \cite{dean2010mapreduce, li2008pfp, riondato2012parma} show that the MapReduce framework is not suitable for frequent itemset mining algorithm like the Apriori algorithm with intensive iterated computation.

\begin{table*}[!htbp]
	\centering
	\small
	\caption{An original sequence database w.r.t. shopping behavior.} 
	\label{table_example1}
	\begin{tabular}{|c|c|c|c|}
		\hline
		\textbf{TID} & \textbf{Time} & \textbf{Customer ID}  & \textbf{Event (products)} \\ \hline
		$t_1$   &   03-06-2017 10:05:30   &   $C_1$  &  \textit{milk}, \textit{bread} 	    \\ 
		$t_2$   &   03-06-2017 10:09:12   &   $C_2$  &  \textit{oatmeal}	    \\ 
		$t_3$   &   03-06-2017 10:21:45   &   $C_3$  &  \textit{milk}, \textit{bread}, \textit{butter}	    \\ 
		$t_4$   &   03-06-2017 11:40:00   &   $C_1$  &  \textit{milk}, \textit{cereal}, \textit{cheese}	    \\ 
		$t_5$   &   03-06-2017 12:55:30   &   $C_3$  &  \textit{cereal}, \textit{oatmeal}	    \\ 
		$t_6$   &   03-06-2017 14:38:58   &   $C_2$  &  \textit{bread}, \textit{milk}, \textit{cheese}, \textit{butter}, \textit{cereal} \\ 
		...   &   ...   &   ...  & ...	    \\ 
		$t_{10}$   &   05-06-2017 15:30:00   &   $C_1$  &  \textit{bread}, \textit{oatmeal}, \textit{butter}	    \\ 						
		\hline
	\end{tabular}
\end{table*}

In the field of data mining, pattern mining has become an important task for a wide range of real-world applications. Pattern mining consists of discovering interesting, useful, and unexpected patterns in databases. This field of research has emerged in the 1990s with the Apriori algorithm \cite{agrawal1994fast} which was proposed by Agrawal and Srikant. It is designed for finding frequent itemsets and then extracting the association rules. Note that frequent itemsets are the groups of items (symbols) frequently appearing together in a database of customer transactions. For example, the pattern/products \{\textit{bread}, \textit{wine}, \textit{cheese}\} can be used to find the shopping behavior of customers for market basket analysis. Some pattern mining techniques, such as frequent itemset mining (FIM) \cite{agrawal1994fast,han2004mining} and association rule mining (ARM) \cite{agrawal1994fast}, are aimed at analyzing data, where the sequential ordering of events is not taken into account. However, the sequence-based database which contains the embedded time-stamp information of event is commonly seen in many real-world applications. A sequence in a sequence database is an ordered list of items, and sequence is everywhere in our daily life. Typical examples include consumers' shopping behavior, Web access logs, DNA sequences in bioinformatics, and so on. We illustrate the sequential data with one case of market basket analysis in detail below. For example, Table \ref{table_example1} is a simple retail store's database which contains customers' shopping records, including transaction ID (TID), occurred time, customer ID, and event (w.r.t. purchase products), and so on. For each customer, his/her total purchase behavior can be considered as a sequence consists of some events which happened at different times. As shown in Table \ref{table_example2} which is translated from Table \ref{table_example1}, the customer $C_1$ w.r.t. sequence ID ($SID$ = 1) has three records ($t_1$, $t_4$, and $t_{10}$) that occurred sequentially.

\begin{table}[!htbp]
	\centering
	\small
	\caption{A translated sequence database from Table 1.} 
	\label{table_example2}
	\begin{tabular}{|c|c|c|}
		\hline
		\textbf{SID} & \textbf{TID}  & \textbf{Event (products)} \\ \hline
		1 &   $t_1$ & 	\textit{milk}, \textit{bread}   \\ 
		1 &   $t_4$ & 	\textit{milk}, \textit{cereal}, \textit{cheese}  \\ 
		1 &   $t_{10}$ & 	\textit{bread}, \textit{oatmeal}, \textit{butter}  \\ 
		2 &   $t_2$ &  	\textit{oatmeal}  \\ 
		2 &   $t_6$ & 	\textit{bread}, \textit{milk}, \textit{cheese}, \textit{butter}, \textit{cereal}  \\  
		3 &   $t_3$ &  	\textit{milk}, \textit{bread}, \textit{butter}	  \\ 
		3 &   $t_5$ & 	\textit{cereal}, \textit{oatmeal}  \\  		
		\hline
	\end{tabular}
\end{table}

To address this issue, another concept of pattern mining named sequential pattern mining (SPM) was first introduced by Agrawal and Srikant in 1995 \cite{agrawal1995mining}. The goal of SPM aims at discovering and analyzing statistically relevant subsequences from sequences of events or items with time constraint. More precisely, it consists of discovering interesting subsequences in a set of sequences, where the interestingness of a subsequence can be measured in terms of various criteria such as its occurrence frequency, length, and profit. Given a user-specified threshold, termed the minimum support (denoted \textit{minsup}), a sequence is said to be a \textit{frequent sequence} or a \textit{sequential pattern} if it occurs more than \textit{minsup} times in the processed database. For instance, consider the database of Table \ref{table_example2}, and assume that the user sets \textit{minsup} = 3 to find all subsequences appearing in at least three sequences. For example, the patterns $\{milk\}$ and  $\{bread, butter\}$ are frequent and have a support of 4 and 3 sequences, respectively. As a fundamental task of pattern mining, sequential pattern mining is used in a wide variety of real-life applications, such as market basket analysis \cite{agrawal1995mining}, Web mining \cite{boggan2007gpus}, bioinformatics \cite{chiu2004efficient}, classification \cite{ayres2002sequential}, finding copy-paste bugs in large-scale software code \cite{chang2008bigtable}.

\begin{figure*}[!htbp]
	\setlength{\abovecaptionskip}{0pt}
	\setlength{\belowcaptionskip}{0pt}	
	\centering
	\includegraphics[width=4.0in]{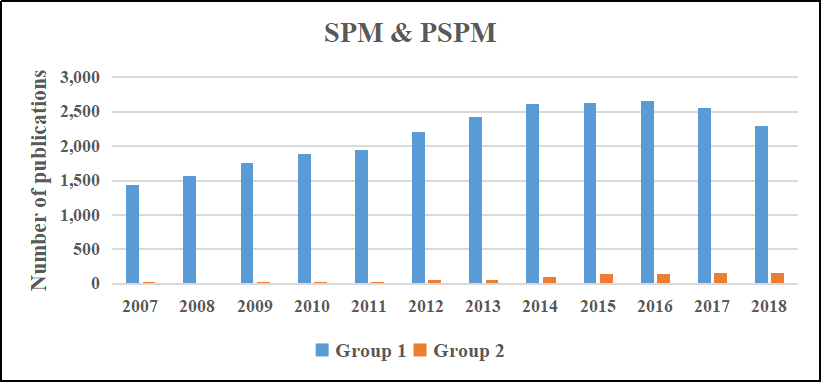}
	\caption{Number of published papers that use ``Parallel Sequential Pattern Mining" in subareas of data mining and big data. These publication statistics are obtained from \textit{Google Scholar}; the search phrase of \textit{Group 1} and \textit{Group 2} is defined as the subfield named with the exact phrase ``parallel sequential pattern" and at least one of parallel or sequential pattern/sequence appearing, e.g., ``parallel sequence", ``parallel sequential pattern".}
	\label{fig:PublicationOfPSPM}
\end{figure*}

In the past two decades, pattern mining (i.e., FIM, ARM and SPM) has been extensively studied and successfully applied in many fields \cite{fournier2017survey0,fournier2017survey}. Meanwhile, to meet the demand of large-scale and high performance computing, as mentioned before, parallel data mining has received considerable attention over the past decades \cite{zaki1999parallel,gan2017data, tsai2014data, tsai2015big}, including parallel frequent itemset mining (PFIM) \cite{apiletti2015pampa, li2008pfp, salah2017highly}, parallel association rule mining (PARM) \cite{riondato2012parma, zaki1997parallel}, parallel sequential pattern mining (PSPM) \cite{beedkar2015lash, zaki2001parallel}, parallel clustering \cite{xu2014efficient, zhao2009parallel}, and so on. Among them, the sequence-based task - PSPM is crucial in a wide range of real-word applications. For example, in Bioinformatics for DNA sequence analysis \cite{chiu2004efficient}, it requires a truly parallel computing on massive large-scale DNA. On one hand, the serial sequential pattern mining is computationally intensive. Although a significant amount of developments have been reported, there is still much room for improvement in its parallel implementation. On the other hand, many applications are time-critical and involve huge volumes of sequential data. Such applications demand more mining power than serial algorithms can provide \cite{cong2005parallel}. Thus, solutions of the sequential pattern mining task that scale are quite important. Up to now, the problem of parallel sequential pattern mining has attracted a lot of attention \cite{beedkar2015lash, zaki2001parallel}. Fig. \ref{fig:PublicationOfPSPM} shows the number of published papers from 2007 to 2017 where \textit{Group 1} denotes the search keywords of ``Sequential Pattern Mining" / ``Sequence Mining", and \textit{Group 2} denotes the search keywords of ``Parallel Sequential Pattern Mining" / ``Parallel Sequence Mining" / MapReduce ``sequential pattern". Fig. \ref{fig:PublicationOfPSPM} outlines a rapid surge of interest in SPM and PSPM in recent years. These results can easily show the trend that mining sequential patterns in a parallel computing environment has emerged as an important issue.

Up to now, several related surveys of serial sequential pattern mining (abbreviated as SPM) have been previously studied \cite{mabroukeh2010taxonomy, mooney2013sequential, zhao2003sequential}. Recently, Fournier-Viger et al. published an up-to-date survey of the current status of serial SPM \cite{fournier2017survey}. There is, unfortunately, no survey of parallel sequential pattern mining methods yet.  In 2014, Anastasiu et al. published an article on the big data frequent pattern mining \cite{anastasiu2014big}, which describes several serial and parallel approaches of scalable frequent itemset mining, sequence mining, and frequent graph mining. However, this article is not specific for parallel sequential pattern mining, only 12 algorithms for PSPM are presented in one section. Many concepts, technologies, big data platforms and tools, domain applications, and the state-of-the-art works of PSPM are not considered and reviewed. Yet, after more than ten years of theoretical development of big data, a significant number of new technologies and applications have appeared in this area. Thus, we  review and summarize current status (i.e., the big data platforms and tools, the new technologies, application, and advanced topics) of PSPM in details.    

The question then posed is: how can one best summarize the related studies in various types of parallel sequential pattern mining in parallel computing environments and make a general taxonomy of them? The methods summarized in this article not only fit for parallel computing \cite{anderson2008general, boggan2007gpus, sousa1998data}, but are also good references for other related work, such as data mining \cite{chen1996data}, big data technologies \cite{khan2014big, labrinidis2012challenges, tsai2014data, tsai2015big, wu2014data}, distributed systems \cite{chang2008bigtable, tanenbaum2007distributed}, and database management \cite{li2014distributed}. Thus, this paper aims at reviewing the current status of parallel sequential pattern mining (PSPM), and  providing in-depth descriptions on a number of representative algorithms. The main contributions of this paper are described below.

\begin{itemize}
	\item We review some related works on serial sequential pattern mining in several categories, including Apriori-based techniques for SPM, pattern growth techniques for SPM, algorithms for SPM with early-pruning, and constraint-based algorithms for SPM. 
	\item We review the state-of-the-art works of parallel sequential pattern mining on distributed environments in recent years. This is a high level survey about parallel techniques for sequential pattern mining in several aspects, including partition-based algorithms for PSPM, Apriori-based PSPM, pattern growth based PSPM, and hybrid algorithms for PSPM. Not only the representative algorithms, but also the advances on latest progresses are reviewed comprehensively. We also point out the key idea, advantages and disadvantages of each PSPM approach.
	\item Some advanced topics for PSPM are reviewed, including PSPM with quantitative / weighted / utility, PSPM from uncertain data and stream data, hardware accelerator for PSPM (i.e., CPU and GPU approaches for parallel sequential pattern mining). 
	\item We further review some well-known open-source software in the fields of serial SPM and parallel SPM, hope that these resources may reduce barriers for future research.
	\item Finally, some challenges and opportunities of parallel sequential pattern mining task for future research are briefly summarized and discussed. 
\end{itemize}

The rest of this paper is organized as follows. Section 2 first introduces the key characteristics about some parallelism methods, then reviews some important features of parallel computing platforms and distributed systems, and respectively summarizes some important parallel computing platforms and tools. The definitions of major concepts used in literature are first introduced in Section 3. Besides, Section 3 also briefly provides the state-of-the-art research efforts SPM. Section 4 highlights and discusses the state-of-the-art research efforts on different approaches, under four taxonomies, for parallel sequential pattern mining. Some advanced topics for PSPM and the related open-source software are further reviewed in Section 5 and Section 6, respectively. Section 7 briefly summarizes some challenges and opportunities in parallel sequential pattern mining. Finally, some conclusions are briefly given in Section 8.

\section{Parallel Computing Framework}
\label{sec:2}

\subsection{Methods of Parallelism}

Zaki has pointed several reasons for designing parallel algorithms to the data mining task \cite{zaki1999parallel,zaki2000parallel}. First, single processors cause the memory and CPU speed limitations, while multiple processors can parallelly reduce them. Second, large amounts of data are already available in parallel sub-databases or they are stored/distributed at multiple sites \cite{zaki1999parallel,zaki2000parallel}. Therefore, it is prohibitively expensive either to collect all data into one site, or to perform the serial mining process on a single computer. For these reasons, tools that can help in parallelization are urgently needed. According to the previous studies \cite{zaki1999parallel,zaki2000parallel}, there are some well-known methods of parallelism: task parallelism (divide and conquer strategy, task queue), data parallelism (record based and attribute based), and hybrid data/task parallelism. Characteristics about these parallelism methods are described below. 

\textbf{(1) Task parallelism}\footnote{\url{http://en.wikipedia.org/wiki/Task_parallelism}} Task parallelism assigns portions of the search space to separate processors. Thus, different tasks running on the same data. There are two types of task parallel approaches: based on divide and conquer strategy and based on a task queue. The former divides the search space and assigns each partition to a specific processor, while the later dynamically assigns small portions of the search space to a processor whenever it becomes available.

\textbf{(2) Data parallelism}\footnote{\url{http://en.wikipedia.org/wiki/Data_parallelism}} 
Data parallelism distributes the dataset over the multiple available processors. The same task parallelly run on different data in distributed environment. In general, data-parallel approaches come in two flavors: record-based data parallelism and attribute-based data parallelism. In \cite{zaki1999parallel,zaki2000parallel}, Zaki had pointed that the records or attribute lists in record-based data parallelism are horizontally partitioned among the processors. In contrast, the attributed in attributed-based data parallelism are divided, thus each processor is responsible for an equal number of attributes.

\textbf{(3) Hybrid data/task parallelism} It is a parallel pipeline of tasks, where each task might be data paralleled, and output of one task is the input to the next one. It means that each task can be run in parallel, throughput impacted by the longest-latency element in the pipeline.

Nowadays, parallel computing is the key to improve the performance of computer programs. Currently, data mining approaches to achieve parallelization and distribution can be classified in terms of five main components \cite{zaki1999parallel,zaki2000parallel}: 
\begin{itemize}
	\item Distributed versus shared memory systems;
	\item Data versus task parallelism; 
	\item Static versus dynamic load balancing; 
	\item Complete versus heuristic candidate generation; 
	\item Horizontal versus vertical data layout. 
\end{itemize}

\subsection{Difference between Parallel Computing Platform and Distributed System}

Unlike traditional centralized systems, a distributed system is defined as: one in which components of networked computers communicate and coordinate their actions only by passing messages \cite{chang2008bigtable, tanenbaum2007distributed}. In other words, a distributed system is a collection of autonomous computing elements that appears to its users as a single coherent system. According to \cite{chang2008bigtable, tanenbaum2007distributed}, there are two aspects of distributed systems: (1) independent computing elements and (2) single system w.r.t. middle-ware. There are some important features of distributed systems, including: 1) concurrency, multi-process and multi-threads concurrently execute and share resources; 2) no global clock, where program coordination depends upon on message passing; and 3) independent failure, wherein some processes’ failure cannot be known by other processes \cite{chang2008bigtable, tanenbaum2007distributed}. There are many types of distributed systems, such as grids \cite{liu2005agent, luo2007distributed}, peer-to-peer (P2P) systems \cite{rao2010optimal}, ad-hoc networks \cite{xue2006optimal}, cloud computing systems \cite{gkatzikis2013migrate}, and online social network systems \cite{jiang2014understanding}. According to a study by Almasi et al., parallel computing\footnote{\url{https://en.wikipedia.org/wiki/Parallel_computing}} is a type of computation in which many calculations or the execution of processes are carried out simultaneously \cite{almasi1988highly}. In general, large problems can often be divided into smaller ones, which can then be solved at the same time. In summary, there are three types of parallelism for parallel computing, including bit-level parallelism, instruction-level parallelism, and task parallelism.

Both distributed mining and parallel mining can speed up the data mining process. However, they have several differences. In 1999, Zaki pointed out that there are some differences between parallel and distributed association rule mining \cite{zaki1999parallel}. For achieving the parallelism of a traditional mining method, the main challenges include: i) synchronization minimization and communication minimization, ii) workload balancing, iii) finding good data layout and data decomposition, and iv) disk I/O minimization (which is especially important for association rule mining) \cite{zaki1999parallel}. In general, the parallel paradigm is hardware or software-distributed shared-memory systems. Thus, one parallel method can be distributed or shared-memory. In distributed data mining (DDM), the same or different mining algorithms are often applied to tackle local data. A DDM algorithm communicates among multiple process units, and then combines the local patterns into the desired global patterns. While a parallel data mining (PDM) algorithm applies the global parallel algorithm on the entire dataset to discover the desired knowledge, it is the same as that found by the traditional data mining algorithm. The accuracy or efficiency of DDM depends on data partitioning, task scheduling, and global synthesizing, and thus it is somewhat difficult to predict \cite{zeng2012distributed}. PDM accuracy may be more guaranteed than that of DDM.

\subsection{Parallel Data Mining Platforms and Tools}

In order to achieve high-performance data mining, some parallel data mining platforms and tools have been further developed in recent years. Many researchers provide different techniques to work on parallel or distributed environments like multi-core computing \cite{ranger2007evaluating,dolbeau2007hmpp}, grid computing \cite{liu2005agent, luo2007distributed}, graphics processing units (GPUs) \cite{anderson2008general, boggan2007gpus}, cloud computing \cite{gkatzikis2013migrate}, Hadoop\footnote{\url{http://hadoop.apache.org}}, etc. In recent years, the focus on computer engineering research shifted to exploit architecture advantages as much as possible, such as shared memory \cite{fournier2013mining}, FPGA \cite{hauck2010reconfigurable}, cluster architecture \cite{almasi1988highly}, or the massive parallelism of GPUs \cite{beedkar2015closing}. Some parallel data mining platforms and tools are first described below.

\textbf{(1) Multi-core computing.} A multi-core processor\footnote{\url{https://en.wikipedia.org/wiki/Multi-core_processor}} includes multiple processing units (called "\textit{cores}") on the same chip. This processor supports multi-core computing, and differs from a super-scalar processor. It can issue multiple instructions per clock cycle from multiple instruction streams \cite{dolbeau2007hmpp}. Recently, a SPM model using multi-core processors is presented in \cite{huynh2017efficient}.

\textbf{(2) Grid computing}. Generally speaking, grid computing\footnote{\url{https://en.wikipedia.org/wiki/Grid_computing}} is a form of parallel computing, and different from conventional computing systems. It makes use of the collection of multiple distributed computer resources to fulfill a common goal. Each node in grid computers performs a different task/application. 

\textbf{(3) Reconfigurable computing with FPGA}. A field-programmable gate array (FPGA) is, in essence, a computer chip that can rewire itself for a given task \cite{hauck2010reconfigurable}. A new computing paradigm, reconfigurable computing, has received wide attention and is the focus of extensive research. Reconfigurable computing uses a FPGA as a co-processor to a general-purpose computer. Thus, high-performance computing can be achieved by using FPGA. Reconfigurable computing with FPGA has a good ability to combine efficiency with flexibility.

\textbf{(4) Graphics processing units (GPUs)}\footnote{\url{https://en.wikipedia.org/wiki/Graphics_processing_unit}}. GPUs have attracted a lot of attention due to their cost-effectiveness and enormous power for massive data parallel computing. At present, a fairly recent trend is to enable general-purpose computing on graphics processing units. GPUs are co-processors that have been heavily optimized for computer graphics processing \cite{anderson2008general, boggan2007gpus}. Recently, some GPU-based tools are developed, such as a parallel graph exploration system on multi-core CPU and GPU \cite{hong2011efficient}, and a novel parallel algorithm on GPUs to accelerate pattern matching \cite{lin2013accelerating}.

\textbf{(5) MapReduce} \cite{dean2010mapreduce}. The MapReduce model was originally proposed by Google. It is a popular parallel programming model that not only simplifies the programming complexity of parallel or distributed computing, but also can achieve better processing and analytics for massive datasets. The MapReduce model \cite{dean2010mapreduce} stores the data in $<$\textit{key, value}$>$ pair and then runs in rounds, which are composed of three consecutive phases: \textit{map}, \textit{shuffle}, and \textit{reduce}. The input and output formats of MapReduce can be expressed as follows: i) \textbf{Mapper}: \textit{$<$key input, value input$>$} to \textit{list$<$key map, value map$>$}; ii) \textbf{Reducer}: \textit{$<$key map, list(values)$>$} to \textit{list$<$key reducer, value reducer$>$}. By using the MapReduce programming model, mining progress can be effectively enhanced. Moreover, I/O cost caused by scanning for massive and high dimensional datasets can be significantly reduced. Up to now, there are many implementations of MapReduce, and MapReduce has become a common powerful tool for parallel or distributed computing under various environments.

\textbf{(6) Spark} \cite{zaharia2012resilient}. Spark is one of the distributed computing framework developed by Berkeley AMPLab. It is a well-known open-source cluster computing system that provides a flexible interface for programming entire clusters with implicit data parallelism and fault-tolerance. Due to the in-memory parallel execution model, all data will be loaded into memory in Spark. This fundamental feature means Spark can significantly speed up computation for iterative algorithms such as the Apriori algorithm and some other machine learning algorithms. In general, Spark \cite{zaharia2012resilient} can be 1-2 orders of magnitude faster than MapReduce \cite{dean2010mapreduce}.

\textbf{(7) PerfExplorer} \cite{huck2005perfexplorer}. It is a computing framework for large-scale parallel data mining. It consists of two main components, the Client and the Server. PerfExplorer has a flexible interface and provides a common, reusable foundation for multiple data analysis, including clustering, dimensionality reduction, coefficient of correlation analysis, and comparative analysis \cite{huck2005perfexplorer}.

\section{State-of-the-art Serial Sequential Pattern Mining} 
\label{sec:background}
Most of the current parallel sequential pattern mining algorithms are extended from the traditional serial SPM algorithms. Therefore, in this section, some preliminaries and the problem statement related to frequent itemset mining (FIM) and sequential pattern mining (SPM) are first introduced. Then, some related works of sequential pattern mining are reviewed in several categories. Finally, the status of SPM is described and briefly summarized.

\subsection{Frequent Itemset Mining VS. Sequential Pattern Mining}
First, let us introduce some preliminaries and the problem statement related to frequent itemset mining (FIM) from transactional databases. Let $I$ = $\{i_1, i_2, $\ldots$, i_n\} $ be a set of items, an \textit{itemset} $X$ = $\{i_1, i_2, $\ldots$, i_k\} $ with $k$ items as a subset of $I$. The length or size of $X$ is denoted as $|X|$, i.e., the number of items in $X$ w.r.t. $k$. Given a transactional database $D$ = $\{t_1, t_2, $\ldots$, t_m\} $, where each transaction $ T_q \in D $ is generally identified by a transaction $id$ (\textit{TID}), and $|D|$ denotes the total number of transactions w.r.t. $m$. The support of $X$ in $D$ is denoted as $sup(X)$, it is the proportion of transactions that contain $X$, i.e., $sup(X)$ = $|{T_q|T_q \in D, X \subseteq T_q}|/|D|$. An itemset is said to be a \textit{frequent itemset} (\textit{FI}) in a database if its support is no less than the user defined minimum support threshold (\textit{minsup}). Therefore, the problem of frequent itemset mining is to discover all itemsets which have a support no less than the defined minimum support threshold, that is $sup(X) \geq  minsup$ \cite{agrawal1994fast}. As the most important mining task for a wide range of real-world applications, frequent itemset mining (FIM) or association rule mining (ARM) has attracted a lot of attention \cite{agrawal1994fast, agrawal1993mining, chen1996data, geng2006interestingness, han2004mining, fournier2017survey0, gan2017data}.

In 1995, Srikant and Agrawal first extended the concept of the frequent itemset mining model to handle sequences \cite{agrawal1995mining}. They first introduced and formulated the sequential pattern mining (SPM) problem. Different from FIM, SPM aims at discovering frequent subsequences as interesting patterns in a sequence database which contains the embedded time-stamp information of event. According to the definitions from Mannila et al. \cite{mannila1997discovery}, we have the associated notations and descriptions of SPM as below. Given an input sequence database \textit{SDB} = $\{S_1, S_2, $\ldots$, S_j\}$, where \textit{SID} is a unique sequence identifier (also called \textit{sequence-id}) and $S_k$ is an input sequence. Thus, \textit{SDB} is a set of tuples (\textit{SID}, $S$), and each sequence $S$ has the following fields: \textit{SID}, event-time, and the items present in the \textit{event}. For example, in Table 1, consider the first sequence $<$$\{a,d\}, \{c\}, \{d,g\}, \{g\}, \{e\}$$>$, it represents five transactions made by a customer at a retail store. Each single letter represents an item (i.e., $\{a\}$, $\{c\}$, $\{d\}$, etc.), and items between curly brackets represent an itemset (i.e., $\{a,d\}$ and $\{d,g\}$). Simply speaking, a \textit{sequence} is a list of temporally ordered itemsets (also called \textit{events}). 

\begin{table}[!htbp]
	\centering
	\small
	\caption{A sequence database.} 
	\label{table_seq1}
	\begin{tabular}{|c|c|c|}
		\hline
		\textbf{SID} & \textbf{Sequence} \\ \hline
		1 & 	$<\{a,d\}, \{c\}, \{d,g\}, \{g\}, \{e\}>$   \\ 
		2 & 	$<\{a\}, \{d\}, \{c,g\}, \{e\}>$  \\ 
		3 & 	$<\{a,b\}, \{c\}, \{f,g\}, \{a,b,g\}, \{e\}>$  \\ 
		4 & 	$<\{b\}, \{c\}, \{d,f\}>$  \\ 
		5 & 	$<\{a,b\}, \{c\}, \{d,f,g\}, \{g\}, \{e\}>$  \\  \hline
	\end{tabular}
\end{table}

A sequence $s_a $ = $<$$A_1, A_2, $\ldots$, A_n$$>$ is called a $k$-\textit{sequence} if it contains $k$ items, or in other words if $k$ = $|A_1|$ + $|A_2|$ + $\ldots$ + $|A_n|$. For example, as shown in Table \ref{table_seq1}, the sequence $<$$\{a,d\}, \{c\}, \{d,g\}, \{g\}, \{e\}$$>$ is a 7-sequence. A sequence $S_\alpha$ = $<$$\alpha_1, \alpha_2, \ldots, \alpha_n $$>$ is called a \textit{sub-sequence} of another sequence $S_\beta$ = $<$$ \beta_1, \beta_2, $\ldots$, \beta_m $$>$ $(n < m)$, and $S_\beta$ is called a \textit{super-sequence} of $S_\alpha$ if there exists an integer $ 1 < i_1< $\ldots$ < i_n < m $ such that $ \beta_{1} \subseteq \beta_{i1}, \ldots, \beta_{in} \subseteq  \beta_{im} $, denoted as $ S_\alpha \subseteq  S_\beta$. A tuple (\textit{SID}, $S$) is said to contain a sequence $ S_\alpha$ if $S$ is a \textit{super-sequence} of $ S_\alpha$, denoted as $ S_\alpha \subseteq S$. As shown in Table \ref{table_seq1}, the sequence $<$$\{a,d\}, \{c\}, \{g\}$$>$ is contained in sequence $<$$\{a,d\}, \{c\}, \{d,g\}, \{g\}, \{e\}$$>$, while the sequence $<$$\{a,d\}, \{c,d\}, \{g\}$$>$ is not. Especially, the sequence $<$$\{a,b\}, \{c\}$$>$ and $<$$\{a,b\}, \{d\}, \{g\}$$>$ are called $sequence$-extensions of $<$$\{a,b\}$$>$, and $<$$\{a,b,d\}$$>$ is called $item$-extension of $<$$\{a,b\}$$>$. The support of a sequence $ S_\alpha$ in a sequence database \textit{SDB} is the number of tuples that contains $ S_\alpha$, and it is denoted as \textit{$sup(S_\alpha)$}. The size of a sequence database \textit{SDB} (denoted as $|SDB|$) corresponds to the total number of sequences (i.e., the number of customers).

In general, sequential pattern mining is more complex and challenging than frequent itemset mining. There are some important reasons for this. Firstly, due to the absence of time constraints in FIM not presenting in SPM, SPM has a potentially huge set of candidate sequences \cite{han2001prefixspan}. The total number of possible sequential patterns is much larger than that of FIM. If all sequences contain exactly one event, SPM is equal to FIM. For example, consider only 100 distinct items for the SPM problem, there are 100 sequences has their length as 1, while the number of sequences with length 2 is $100 \times 100 \times (100 \times 99/2)$ = 14,950, and the number of sequences with length 3 is $2^{100} - 1 \approx 10^{30}$. Secondly, there are further difficulties mining longer sequential patterns \cite{han2001prefixspan}, such as DNA sequence analysis or stock sequence analytics/prediction.

\subsection{State-of-the-art Serial Sequential Pattern Mining}

Through 20 years of study and development, many techniques and approaches have been proposed for mining sequential patterns in a wide range of real-word applications \cite{fournier2017survey}, such as Web mining \cite{boggan2007gpus}, classification \cite{ayres2002sequential}, and mining motifs from biological sequences \cite{chiu2004efficient}. Some well-known serial algorithms for sequential pattern mining, such as AprioriAll \cite{agrawal1995mining}, GSP \cite{srikant1996mining}, BIDE \cite{wang2004bide}, CloSpan \cite{yan2003clospan}, FreeSpan \cite{han2000freespan}, PrefixSpan \cite{han2001prefixspan}, SPADE \cite{zaki2001spade}, etc., have been proposed. Since many parallel SPM algorithms are extended by the traditional serial SPM approaches, it is quite necessary for us to have an understanding of the systematic characteristics of traditional serial sequential pattern mining approaches. Thus, the state-of-the-art of serial SPM are reviewed below. Based on the different mining mechanisms, some well-known serial SPM approaches are divided into several categories, including Apriori-based techniques, pattern growth techniques, algorithms with early-pruning, and constraint-based algorithms. Details are summarized in Tables \ref{table_1}, \ref{table_2}, \ref{table_3}, and \ref{table_4}, respectively.

\begin{table*}[!htbp]
	\centering
	\small
	\caption{Apriori-based serial algorithms for sequential pattern mining.} 
	\label{table_1}
	\newcommand{\tl}[1]{\multicolumn{1}{l}{#1}}
	\begin{tabular}{|c|l|l|l|} 
		\hline
		\multicolumn{1}{|c|}{\textbf{Name}} & \multicolumn{1}{|c|}{\textbf{Description}} &  \multicolumn{1}{|c|}{\textbf{Year}} \\ \hline		
		
		Apriori (All, Some,         & 	The first algorithm for sequential pattern mining.  & 1995 \\ 
		 Dynamic Some) \cite{agrawal1995mining} &      &   \\    \hline
		
		GSP \cite{srikant1996mining} & 	Generalized Sequential Patterns (max/min gap,  & 1996 \\
		 &  window, and taxonomies). & \\ \hline
		
		PSP \cite{masseglia2000efficient} & 	Retrieval optimizations and more efficient than GSP.  & 1998 \\ \hline
		
		SPADE \cite{zaki2001spade} & 	Sequential PAttern Discovery using Equivalence classes.  & 2001 \\ \hline
		
		SPAM \cite{ayres2002sequential} & 	Sequential PAttern Mining with Bitmap representation.  & 2002 \\ \hline		
		
		LAPIN \cite{yang2007lapin} & 	SPM with LAst Position Induction, which is categorized & 2004 \\
		& into two classes, LAPIN-LCI and LAPIN-Suffix.  &  \\ \hline
		
		LAPIN-SPAM \cite{yang2005lapin} & 	LAst Position INduction Sequential PAttern Mining.  & 2005 \\ \hline
	\end{tabular}
\end{table*}

\begin{table*}[!htbp]
	\centering
	\small
	\caption{Pattern growth algorithms for sequential pattern mining.}
	\label{table_2}
	\newcommand{\tl}[1]{\multicolumn{1}{l}{#1}} 
	\begin{tabular}{|c|l|l|l|} 
		\hline
		\multicolumn{1}{|c|}{\textbf{Name}} & \multicolumn{1}{|c|}{\textbf{Description}} & \multicolumn{1}{|c|}{\textbf{Year}} \\ \hline
		
		FreeSpan \cite{han2000freespan} & 	FREquEnt pattern-projected Sequential Pattern mining.  & 2000 \\ \hline
		
		WAP-mine \cite{pei2000mining} & 	SPM with suffix growth.  & 2000 \\ \hline
		
		PrefixSpan \cite{han2001prefixspan} & 	PREFIX-projected Sequential PAtterN mining.  & 2001 \\ \hline
		
		LPMiner \cite{seno2001lpminer} & 	Sequential pattern mining with Length decreasing suPport.  & 2001 \\ \hline

		SLPMiner \cite{seno2002slpminer} & 	Sequential pattern mining with Length decreasing suPport.  & 2002 \\ \hline

		FS-Miner \cite{el2004fs} & 	SPM with suffix growth.  & 2004 \\ \hline
		
		LAPIN-Suffix \cite{yang2007lapin} & 	SPM with suffix growth.  & 2004 \\ \hline
		
		PLWAP \cite{ezeife2005plwap}  & 	SPM with prefix growth.  & 2005 \\ \hline
	\end{tabular}
\end{table*}

\textbf{Apriori-Based Techniques.} In 1995, Agrawal and Srikant first introduced an new approach named AprioriAll to discover the sequential patterns from a set of sequences whithin the embedded timestamp information \cite{agrawal1995mining}. The AprioriAll algorithm is based on Apriori and relies on the Apriori property. That is ``all nonempty subsets of a frequent itemset must also be frequent'' \cite{agrawal1994fast}. Therefore, AprioriAll holds the \textit{downward closed property} (also called \textit{anti-monotonic}), and utilizes this anti-monotonic to prune the search space. In general, two properties are often utilized in sequential pattern mining to speed up computation: 
\begin{itemize}
	\item \textbf{\textit{Subsequence infrequency based pruning}}: any supersequence of an infrequent sequence is not frequent, thus it can be pruned from the set of candidates \cite{agrawal1995mining}. If $S_a$ is a subsequence of $S_b $, then $sup(S_b)$ is at most as large as $sup(S_a)$. \textit{Monotonicity}: If $S_a$ is not frequent, then it is impossible that $S_b$ is frequent. \textit{Pruning}: If we know that $S_a$ is not frequent, we do not have to evaluate any supersequence of $S_a$. For example, if sequence $<$$a$$>$ occurs only 10 times, then $<$$a, b$$>$ can occur at most 10 times.
	\item \textbf{\textit{Supersequence frequency based pruning}}: any subsequence of a frequent sequence is also frequent, thus it can be safely pruned from the set of candidates \cite{bayardo1998efficiently}. For example, if sequence $<$$a, b, d$$>$ is frequent and occurs 8 times, then any its subsequence ($<$$a$$>$, $<$$b$$>$, $<$$d$$>$, $<$$a, b$$>$, $<$$a, d$$>$ and $<$$b, b$$>$) is frequent and occurs at least 8 times.
\end{itemize}

Srikant then proposed GSP, which is similar to AprioriAll in execution process, but adopts several technologies including time constraints, sliding time windows, and taxonomies \cite{srikant1996mining}. GSP uses a multiple-pass, candidate generation-and-test method to find sequential patterns. It adopts the Apriori property that all sub-patterns of a frequent pattern must be frequent. Thus, GSP can greatly improve performance over AprioriAll. PSP \cite{masseglia2000efficient} resumes the general principles of GSP but it utilizes a different intermediary data structure, making it more efficient than GSP. AprioriAll \cite{agrawal1995mining} AprioriSome \cite{agrawal1995mining}, DynamicSome \cite{agrawal1995mining} and GSP \cite{srikant1996mining} all use the breadth-first search (BFS) to mine sequential patterns with a hierarchical structure. The SPADE algorithm, which uses equivalence classes to discover the sequential patterns, is an Apriori-based hybrid miner, it can be either breadth-first or depth-first \cite{zaki2001spade}. SPADE exploits sequential patterns by utilizing a vertical id-list database format and a vertical lattice structure. SPAM is similar to SPADE, but SPAM uses bitmap representation and bitwise operations rather than regular and temporal joins. Yang et al. then proposed a novel algorithm LAPIN for SPM with last position induction (LAPIN-LCI and LAPIN-Suffix) \cite{yang2007lapin}. They further developed the LAPIN-SPAM \cite{yang2005lapin} algorithm by combining LAPIN and SPAM. Note that for SPAM, LAPIN, LAPIN-SPAM and SPADE, these algorithms use the property that the support of a super-patterns is always less than or equal to the support of its support patterns, it is different from the Apriori property used in GSP. Most of the above Apriori-based algorithms consist of three features. 
\begin{itemize}
	\item  \textbf{Breadth-first search}: Apriori-based algorithms commonly use the breadth-first (w.r.t. level-wise) search manner. They construct all $k$-sequences together in each $k$-th iteration while exploring the search space. 
	\item \textbf{Generate-and-test}: This feature is introduced by Agrawal et al. in the Apriori algorithm \cite{agrawal1994fast}. It is used in the early sequential pattern mining algorithms, such as AprioriAll \cite{agrawal1995mining}, AprioriSome \cite{agrawal1995mining}, Dynamic Some \cite{agrawal1995mining}. 
	\item \textbf{Multiple database scans}: These algorithms need to scan the original database many times to determine whether a longer generated sequences is frequent. They suffer from the drawbacks of the candidate generation-and-test paradigm. In other words, they may generate a huge number of candidate sequences that do not appear in the input database and need to scan the original database many times. 
\end{itemize}

\begin{table*}[!htbp]
	\centering
	\small
	\caption{Algorithms for sequential pattern mining with early-pruning.}
	\label{table_3}
	\newcommand{\tl}[1]{\multicolumn{1}{l}{#1}}  
	\begin{tabular}{|c|l|l|l|} 
		\hline
		\multicolumn{1}{|c|}{\textbf{Name}} & \multicolumn{1}{|c|}{\textbf{Description}} &  \multicolumn{1}{|c|}{\textbf{Year}} \\ \hline
		
		HVSM \cite{song2005hvsm} & 	Bitwise operation and position induction.  & 2005 \\ \hline
		
		LAPIN-SPAM \cite{yang2005lapin} & 	Bitwise operation.  & 2005  \\ \hline
		
		PLWAP \cite{ezeife2005plwap}  & 	Position induction.  & 2005 \\ \hline
		
		LAPIN \cite{yang2007lapin} &   Position induction.  & 2007 \\ \hline

		DISC-all \cite{chiu2004efficient} & Position induction and prefix growth.  & 2007 \\ \hline

		UDDAG \cite{chen2010updown} & 	Up-Down Directed Acyclic Graph for sequential pattern mining.  & 2010 \\ \hline
	\end{tabular}
\end{table*}

\textbf{Pattern-Growth Techniques.} In 2000, a pattern-projected sequential pattern mining algorithm named FreeSpan was introduced by Han et al. \cite{han2000freespan}. By using a frequent item list (f-list) and S-Matrix, it obtains all frequent sequences based on so-called projected pattern growth. Pattern-growth SPM algorithms can avoid recursively scanning the database to find frequent patterns. The reason is that they only consider the patterns actually appearing in the database. To reduce time-consuming database scans, pattern-growth SPM algorithm introduces a novel concept called projected database \cite{han2000freespan,anderson2008general}. The projected database can significantly reduce the size of databases as frequent patterns are considered by the depth-first search.

The WAP-mine algorithm, which utilizes a WAP-tree, was proposed by Pei et al. \cite{pei2000mining}. In a WAP-tree, each node is assigned a binary code to determine which sequences are the suffix sequences of the last event, and then to find the next prefix for a discovered suffix. Thus, PLWAP does not have to reconstruct intermediate WAP-trees. Han et al. then further proposed the most representative algorithm PrefixSpan \cite{han2001prefixspan}. It tests only the prefix subsequences, and then projects their corresponding postfix subsequences into the projected sub-databases (A projected database is the set of suffixes w.r.t. a given prefix sequence). By recursively exploring only local frequent sequences, sequential patterns can be recursively grown in each projected subdatabase. Thus, PrefixSpan exhibits better performance than both GSP \cite{srikant1996mining} and FreeSpan \cite{han2000freespan}. However, when dealing with large dense databases that have large itemsets, it is worse than that of SPADE \cite{zaki2001spade}. As shown in Table \ref{table_2}, some other pattern growth algorithms for sequential pattern mining have been developed, such as LPMiner \cite{seno2001lpminer}, SLPMiner \cite{seno2002slpminer}, WAP-mine \cite{pei2000mining}, and FS-Miner \cite{el2004fs}. With consideration of length decreasing support, the LPMiner \cite{seno2001lpminer} and SLPMiner \cite{seno2002slpminer} algorithms are introduced. In 2005, Ezeife et al. proposed a pattern-growth and early-pruning hybrid method named PLWAP \cite{ezeife2005plwap}. PLWAP utilizes a binary code assignment method to construct a preordered, linked, position-coded Web access pattern tree (WAP-tree for short). Compared to the WAP algorithm, PLWAP can significantly reduce runtime and provide a position code mechanism. Inspired by FP-tree \cite{han2004mining} and the projection technique for quickly mining sequential patterns \cite{anderson2008general}, a tree algorithm named FS-Miner (Frequent Sequence Miner) is developed \cite{el2004fs}. FS-Miner resembles WAP-mine \cite{pei2000mining}. Moreover, it supports incremental mining and interactive mining \cite{fournier2016spmf}.

\begin{table*}[!htbp]
	\centering
	\small
	\caption{Constraint-based algorithms for sequential pattern mining.} 
	\label{table_4}
	\newcommand{\tl}[1]{\multicolumn{1}{l}{#1}}   
	\begin{tabular}{|c|l|l|l|l|l|} 
		\hline
		\multicolumn{1}{|c|}{\textbf{Name}} & \multicolumn{1}{|c|}{\textbf{Description}} & \multicolumn{1}{|c|}{\textbf{Constraint}} & \multicolumn{1}{|c|}{\textbf{Character}} & \multicolumn{1}{|c|}{\textbf{Year}} \\ \hline

		CloSpan \cite{yan2003clospan} & 	\multirow{2}{4.5cm}{The first algorithm for mining closed sequential patterns.}   & \multirow{2}{2cm}{$\bullet$ Closed} & 	\multirow{2}{5cm}{$\bullet$ Based on the GSP algorithm.}  & 2003  \\ 
		& &  & &   \\ \hline
		
		BIDE \cite{wang2004bide} & 	\multirow{2}{4.5cm}{Mining of frequent closed sequences by using BI-Directional Extension, which is a sequence closure checking scheme.}  &  \multirow{2}{2cm}{$\bullet$ Closed} & 	\multirow{2}{5cm}{$\bullet$ It executes some checking steps in the original database that avoid maintaining the sequence tree in memory.}  & 2004  \\
		& &  & &   \\   
		& &  & &   \\  
		& &  & &   \\ \hline
		
		MSPX \cite{luo2005efficient} &     \multirow{2}{4.5cm}{MSPX mines maximal sequential patterns with a bottom-up search and uses multiple samples.}  & \multirow{2}{2cm}{$\bullet$ Maximal} & 	\multirow{2}{5cm}{$\bullet$ The sampling technique is used at each pass to distinguish potentially infrequent candidates.}  & 2005  \\ 
		& &  & &   \\  					
		& &  & &   \\ \hline
		
		SQUIRE \cite{kim2007squire} & 	\multirow{2}{4.5cm}{Sequential pattern mining with quantities, Apriori-QSP and PrefixSpan-QSP.}  & \multirow{2}{2cm}{$\bullet$ Pattern with quantities} & 	\multirow{2}{5cm}{$\bullet$ Apriori-based and PrefixSpan-based.}  & 2007  \\ 
		& &  & &   \\  	
		& &  & &   \\ \hline

		ClaSP \cite{gomariz2013clasp} &   \multirow{2}{4.5cm}{Mining frequent closed sequential patterns by using several search space pruning methods together with a vertical database layout.} & \multirow{2}{2cm}{$\bullet$ Closed} & 	\multirow{2}{5cm}{$\bullet$ Inspired on the SPADE and CloSpan algorithms.}  & 2013  \\ 
		& &  & &   \\  
		& &  & &   \\  
		& &  & &   \\ \hline

		MaxSP \cite{fournier2013mining}  & 	\multirow{2}{4.5cm}{Maximal Sequential Pattern miner without candidate maintenance.}  & \multirow{2}{2cm}{$\bullet$ Maximal} & 	\multirow{2}{5cm}{$\bullet$ It neither produces redundant candidates nor stores intermediate candidates in main memory.}  & 2013  \\
		& &  & &   \\   
		& &  & &   \\ \hline

		VMSP \cite{fournier2014vmsp}  & 	\multirow{2}{4.5cm}{The first vertical mining algorithm for Vertical mining of Maximal Sequential Patterns.}  & \multirow{2}{2cm}{$\bullet$ Maximal} & 	\multirow{2}{5cm}{$\bullet$ Uses a vertical representation to do a depth-first exploration of the search space.}  & 2014  \\
		& &  & &   \\     
		& &  & &   \\ \hline
		
		CloFAST \cite{fumarola2016clofast} & 	\multirow{2}{4.5cm}{A fast algorithm for mining closed sequential patterns using id-list.}  & \multirow{2}{2cm}{$\bullet$ Closed} & 	\multirow{2}{5cm}{$\bullet$ It combines	a new data representation of the dataset, based on sparse and vertical id-list.}  & 2016  \\ 
		& &  & &   \\   
		& &  & &   \\ \hline
	\end{tabular}
\end{table*}

\textbf{Early-Pruning Techniques.} Although the projection-based approaches (i.e., FreeSpan, PrefixSpan) can achieve a significant improvement over Apriori-based approaches, the projection mechanism still suffers from some drawbacks. The major cost of PrefixSpan is caused by constructing projected databases. Therefore, some hybrid algorithms with early-pruning strategy are developed, as shown in Table \ref{table_3}. Following in the footsteps of SPAM, a first-Horizontal-last-Vertical scanning database Sequential pattern Mining algorithm (named HVSM for short) \cite{song2005hvsm} is developed using bitmap representation. Instead of using candidate generate-and-test or the tree projection, the LAPIN algorithm \cite{yang2007lapin} uses an item-last-position list and a  position set of prefix border. Based on the anti-monotone property which can prune infrequent sequences, Direct Sequence Comparison (DISC-all) uses other sequences of the same length to prune infrequent sequences \cite{chiu2004efficient}. The DISC-all respectively employs different orderings (lexicographical ordering and temporal ordering) to compare sequences of the same length. With the up-to-down directed acyclic graph, the UDDAG algorithm \cite{chen2010updown} uses the prefixes and suffixes to detect the pattern. It obtains faster pattern growth because of fewer levels of database projection compared to other approaches, and allows bi-directional pattern growth along both ends of detected patterns.

\textbf{Constraint-based Techniques.} Different from the traditional SPM approaches, the interesting issue of constraint-based SPM has been widely studied, including quantitative sequences, maximal sequential pattern, closed sequential patterns, sequential patterns with gap constraint, top-$k$ sequential patterns, etc. In other perspectives, SPM with regular expression based constraints \cite{trasarti2008sequence}, SPIRIT algorithm \cite{garofalakis1999spirit}, and SPM with declarative constraints \cite{beedkar2016desq} have been extensively studied.

Some constraint-based SPM algorithms are described in Table \ref{table_4}. In order to handle the condense representation of an explosive number of frequent sequential patterns, the issue of mining maximal sequential patterns is first developed. The MSPX mines maximal sequential patterns with a bottom-up search and uses multiple samples \cite{luo2005efficient}. Unlike the traditional single-sample techniques used in FIM algorithms, MSPX uses multiple samples at each pass to distinguish potentially infrequent candidates. Then, an efficient algorithm called MaxSP (Maximal Sequential Pattern miner without candidate maintenance) was proposed \cite{fournier2013mining}. The maximal representation may cause the information loss of support, thus another condense representation named closed pattern was introduced \cite{yan2003clospan}. Up to now, some algorithms for mining closed sequential patterns have been proposed, such as CloSpan \cite{yan2003clospan}, BIDE \cite{wang2004bide}, ClaSP \cite{gomariz2013clasp},  CloFS-DBV \cite{tran2015combination}, and CloFAST \cite{fumarola2016clofast}. CloSpan \cite{yan2003clospan} adopts the candidate maintenance-and-test method to prune the search space, and it may perform worse while database having long sequences or the support threshold is very low. A novel closure checking scheme, called bi-directional extension, is developed in BIDE \cite{wang2004bide}. BIDE mines closed sequential patterns without candidate maintenance, it is more efficient than CloSpan. Gomariz et al. then introduced the ClaSP algorithm by using several efficient search space pruning methods together with a vertical database layout \cite{gomariz2013clasp}. Recently, a more fast algorithm called CloFAST was proposed for mining closed sequential patterns using sparse and vertical id-lists \cite{fumarola2016clofast}. CloFAST does not build pseudo-projected databases and does not need to scan them. It is more efficient than the previous approaches, including CloSpan, BIDE, and ClaSP.

\textbf{Discussions.} Each method of serial sequential pattern mining algorithm has advantages and disadvantages. Besides, there are also many different definitions of categories for SPM algorithms. A more comprehensive discussion of these methods has been given in \cite{mabroukeh2010taxonomy, mooney2013sequential, zhao2003sequential}, and an up-to-date survey of the current status of sequent pattern minng can be referred to \cite{fournier2017survey}.

\section{State-of-the-art Parallel Sequential Pattern Mining} 

In this section, we first briefly overview the current status of parallel frequent itemset mining. We also cover some special categories of parallel sequential pattern mining algorithms, i.e., partition-based algorithms for PSPM, Apriori-based algorithms for PSPM, pattern growth algorithms for PSPM, and hybrid algorithms for PSPM. In most PSPM algorithms, they are hybrid inherently.

\subsection{Parallel Frequent Itemset Mining} 

As mentioned previously, sequential pattern mining is highly related to frequent itemset mining, and SPM is more complicated than FIM. Since some developments of SPM are inspired by many technologies of FIM, it is needed to have an overview of the current development in parallel frequent itemset mining (PFIM). The problem of frequent itemset mining in parallel environments has been extensively studied so far, and a number of approaches have been explored to address this problem, such as PEAR \cite{mueller1998fast}, PPAR \cite{mueller1998fast}, PDM \cite{park1995efficient}, ParEclat \cite{zaki1997parallel}, PaMPa-HD \cite{apiletti2015pampa}, PHIKS \cite{salah2017highly}, etc. In 1995, Mueller first proposed two parallel algorithms, Parallel Efficient Association Rules (PEAR) \cite{mueller1998fast} and Parallel Partition Association Rules (PPAR) \cite{mueller1998fast}. Cheung et al. also proposed an algorithm named Parallel Data Mining (PDM) to parallel mining of association rules \cite{park1995efficient}, and the Fast Distributed Mining (FDM) algorithm for distributed databases \cite{cheung1996fast}. An adaptation of the FP-Growth algorithm to MapReduce \cite{dean2010mapreduce} called PFP (parallel FP-tree) is presented in \cite{li2008pfp}. PFP is a parallel form of the classical FP-Growth, it splits a large-scale mining task into independent and parallel tasks. By extending the vertical mining approach Eclat \cite{zaki1997new}, the parallel-based Eclat (ParEclat) \cite{zaki1997parallel} and the distributed Eclat (Dist-Eclat) \cite{moens2013frequent} were developed, respectively. Research efforts have already been made to improve Apriori-based and traditional FIM algorithms and to convert them into parallel versions, mostly under the MapReduce \cite{dean2010mapreduce} or Spark \cite{zaharia2012resilient} environment. Some results of these efforts are PARMA \cite{riondato2012parma}, a parallel frequent itemset mining algorithm with Spark (R-Apriori) \cite{rathee2015r}, PaMPa-HD \cite{apiletti2015pampa}, and PHIKS \cite{salah2017highly}. PARMA \cite{riondato2012parma} is a parallel algorithm for the MapReduce framework, which mines approximate association rules instead of exact ones.

\textbf{Discussions.} The above parallel algorithms are used in frequent itemset mining. Up to now, large numbers of studies of parallel frequent itemset mining have been extensively studied  compared to parallel sequential pattern mining. The development of PFIM is still an active area in research, where the up-to-date advances of PFIM can be referred to \cite{anastasiu2014big,gan2017data}. As mentioned before, efficient mining of frequent sequences is more challenging than FIM. Currently, a large number of algorithms have been extensively developed for SPM, while few of them are sufficiently scalable to handle large-scale datasets. 

\subsection{Partition-based Algorithms for PSPM}

In addition to parallel frequent itemset mining,  a variety of mining algorithms has been developed for parallel sequential pattern mining (PSPM). Although PSPM is comparatively more complicate and challenging than that of PFIM, research in the area of PSPM has been active for some time. Recently, to improve the performance, effectiveness and scalability issues, parallel sequential pattern mining has received much attention. 

Firstly, the partition-based parallel methods for SPM are introduced and reviewed. The central idea behind partition-based algorithms is described below. Here the partition mainly in terms of the computation partition and data partition. For example, the database itself can be shared, partially or totally replicated, or partitioned among the available distributed nodes \cite{zaki2000parallel}. Notice that the partition can be achieved by using round-robin, hash, or range scheduling. As shown in Table \ref{table_disPartition}, some partition-based parallel methods for SPM have been developed. A comprehensive working example of partition-based sequential pattern mining algorithm with task parallelism is illustrated below, to show how the problem of parallel sequential pattern mining is solved. It could help to understand the idea of parallel sequential pattern mining.

\begin{figure}[!htbp]
	\setlength{\abovecaptionskip}{0pt}
	\setlength{\belowcaptionskip}{0pt}	
	\centering
	\includegraphics[width=4.0in]{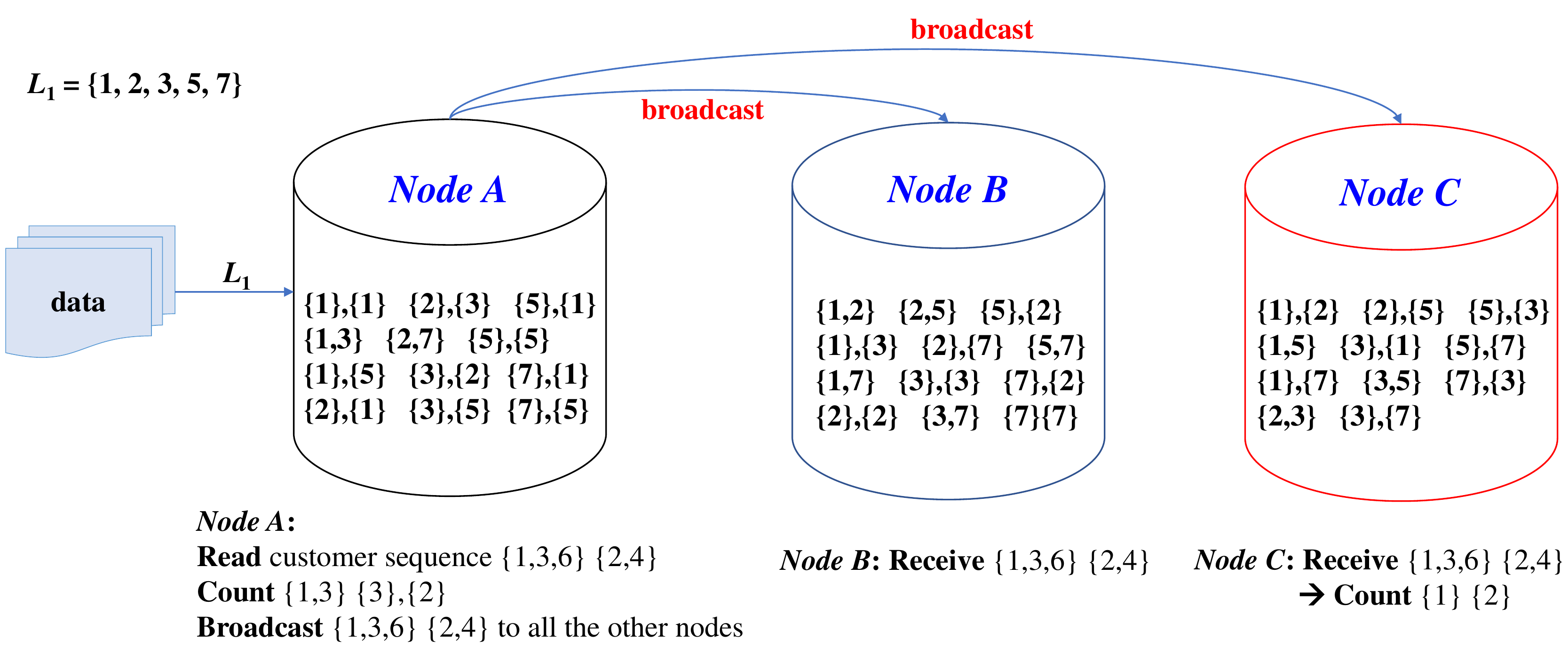}
	\caption{An illustrated example of the SPSPM algorithm \cite{shintani1998mining}.}
	\label{fig:ExampleOfSPSPM}
\end{figure}

\begin{table*}[!htbp]
	\centering
	\small
	\caption{Partition-based algorithms for parallel sequential pattern mining.}
	
	\label{table_disPartition}
	\newcommand{\tl}[1]{\multicolumn{1}{l}{#1}}    
	\begin{tabular}{|c|l|l|l|l|l|} 
		\hline
		\multicolumn{1}{|c|}{\textbf{Name}} & \multicolumn{1}{|c|}{\textbf{Description}} & \multicolumn{1}{|c|}{\textbf{Pros.}} & \multicolumn{1}{|c|}{\textbf{Cons.}} & \multicolumn{1}{|c|}{\textbf{Year}} \\ \hline
		
		HPSPM \cite{shintani1998mining} & 	\multirow{2}{3.5cm}{Hash Partitioned Sequential Pattern Mining, task parallelism.} & 	\multirow{2}{4cm}{$\bullet$ It can considerably reduces the broadcasting of sequences.} & 	\multirow{2}{4cm}{$\bullet$ It is based on GSP and has a level-wise candidate generation-and-test style.}  & 1998  \\ 
		& &  & &   \\ 
		& &  & &   \\ \hline
		
		SPSPM \cite{shintani1998mining}  & 	\multirow{2}{3cm}{Simply Partitioned Sequential Pattern Mining, task parallelism.}  & \multirow{2}{4cm}{$\bullet$ The candidate sets assigned to the nodes have almost the same size.\\
			$\bullet$ They are disjoint from each other.} & 	\multirow{2}{4cm}{$\bullet$ It requires a very large communication overhead.}  & 1998  \\ 
		& &  & &   \\ 
		& &  & &   \\ 
		& &  & &   \\ 
		& &  & &   \\ \hline
		
		NPSPM \cite{shintani1998mining} & 	\multirow{2}{3cm}{Non-Partitioned Sequential Pattern Mining, data parallelism.}  & \multirow{2}{4cm}{$\bullet$  Mining sequential patterns without partition.} & 	\multirow{2}{4cm}{$\bullet$ NPSPM is based on GSP and performed in a bottom-up, level-wise manner with multiple data scans.}  & 1998  \\ 
		& &  & &   \\ 
		& &  & &   \\ 
		& &  & &   \\ \hline

		EVE \cite{joshi1999parallel} & 	\multirow{2}{3cm}{EVEnt distribution.}  & \multirow{2}{4cm}{$\bullet$ Assume that the entire candidate sequences can be replicated and fit in the overall memory of a process.} & 	\multirow{2}{4cm}{$\bullet$ The I/O overhead is impractical in the big data context.}  & 1999  \\ 
		& &  & &   \\ 
		& &  & &   \\ 
		& &  & &   \\ \hline

		EVECAN \cite{joshi1999parallel} & 	\multirow{2}{3cm}{EVEnt and CANdidate distribution.}  & \multirow{2}{4cm}{$\bullet$ Rotates the smaller of two sets among the processes with a round-robin fashion.} & 	\multirow{2}{4cm}{$\bullet$ The I/O overhead is impractical in the Big Data context.}  & 1999  \\ 
		& &  & &   \\ 
		& &  & &   \\ \hline

		\multirow{2}{2cm}{DPF, STPF \\
			and DTPF \cite{guralnik2004parallel}} & 	\multirow{2}{3cm}{Data Parallel Formulation,\\
			Static Task-Parallel Formulation, and\\
			Dynamic Task-Parallel Formulation}  & \multirow{2}{4cm}{$\bullet$ It can achieve balanced workloads using a bi-level breadth-first manner in  the lexicographic sequence tree.} & 	\multirow{2}{4cm}{$\bullet$ At each iteration, it has to partition the tree and perform multiple scans of the local database.\\ $\bullet$ This added I/O overhead is impractical when mining big data.}  & 2004  \\ 
		& &  & &   \\ 
		& &  & &   \\ 
		& &  & &   \\ 
		& &  & &   \\ 
		& &  & &   \\ 
		& &  & &   \\ \hline
	\end{tabular}
\end{table*}

Note that Fig. \ref{fig:ExampleOfSPSPM} is an illustrated example of the SPSPM algorithm (Simply Partitioned Sequential Pattern Mining with task parallelism) \cite{shintani1998mining}. Assuming there are 3 nodes in the mining system, denoted as \textit{Node A}, \textit{Node B} and \textit{Node C}, respectively. Each node reads the sequence database $D_k$ from its local disk, and let the large/frequent items be $L_1 = \{1,2,3,5,7\}$. Since SPSPM partitions/distributes the candidate sequences in a round-robin manner, the candidate sequences are assigned shown in Fig. \ref{fig:ExampleOfSPSPM}. Suppose \textit{Node A} reads the sequence $s = \{1,3,6\}\{2,4\}$. Then \textit{Node A} broadcasts $s$ stored in its local disk to all the other nodes, and increases the count-support of $\{1,3\},\{3\}\{2\}$. \textit{Node B} and \textit{Node C} respectively receive the sequence $\{1,3,6\},\{2,4\}$ from \textit{Node A}. Then \textit{Node C} increases the support-count of $\{1,2\}$.

\textbf{NPSPM, SPSPM and HPSPM.} In 1998, Shintani and Kitsuregawa firstly partitioned the input sequences for SPM, and yet they assumed that the entire candidate sets could be replicated and would fit in the overall memory (random access memory and hard drive) of a process \cite{shintani1998mining}. Based on the classical GSP approach \cite{srikant1996mining}, they proposed three parallel approaches \cite{shintani1998mining}: NPSPM (Non-Partitioned Sequential Pattern Mining), SPSPM (Simply Partitioned Sequential Pattern Mining), and HPSPM (Hash Partitioned Sequential Pattern Mining). These three variants perform as a classical level-wise candidate generation-and-test style. HPSPM is similar to SPSPM but utilizes more intelligent strategies. HPSPM uses the hash mechanism to assign the input sequences and candidate sequences to specific processes \cite{shintani1998mining}. Although HPSPM has the best performance among the three variants, it also suffers, however, several limitations. For exampe, when the object time-lines are very large, or when the counting method for generalized SPM is used, HPSPM is very time-consuming and memory cost.

\textbf{EVE \& EVECAN.} Similar assumptions of NPSPM were made in EVE \cite{joshi1999parallel} and EVECAN \cite{joshi1999parallel}. EVECAN (EVEnt and CANdidate distribution) partitions the input sequentail data similar to EVE (EVEnt distribution), and also partitions the candidate sequences. In EVECAN, both the candidate generation phase and the counting phase are parallelized. The former uses a distributed hash table mechanism, whereas the later adopts an approach similar to Intelligent Data Distribution (IDD), which is usually used for non-sequential associations. EVECAN shards both input sequences and candidate sequences, and rotates the smaller of the two sequence sets among the processes, in round-robin fashion. Note that the above parallel formulations for SPM only toward the shared-memory architecture.

\textbf{DPF, STPF and DTPF}. Guralnik and Karypis developed three tree-projection-based parallel sequence mining algorithm \cite{guralnik2004parallel}, named data parallel formulation (DPF), static task-parallel formulation (STPF), and dynamic task-parallel formulation (DTPF), for distributed-memory architectures. These projection algorithms are parallelized using the data parallel formulation (DPF) and the task parallel formulation (TPF). They are intrinsically similar to the PrefixSpan algorithm \cite{han2001prefixspan}, grow the lexicographic sequence tree in a bi-level breadth-first manner. Two partitioning methods were proposed: TPF-BP (TPF based on a bin-packing algorithm) and TPF-GP (TPF based on a minimum-cut bipartite graph partitioning algorithm), both TPF-BP and TPF-GP can achieve balanced workloads. It was shown that these three algorithms are able to achieve good speedups when the number of processors increases. However, STPF may suffer some load imbalance problem while the number of processes increases, and DPF has not only to partition the tree, but also needs to perform multiple scans of the local database at each iteration. The added I/O overhead makes DPF impractical for big data analytics. In addition, DPF utilizes a dynamic load-balancing strategy to allow an idle processor to join the busy processors. However, this dynamic strategy requires much more inter-processor communication than the selective sampling approach used in the Par-CSP algorithm \cite{cong2005parallel}. Moreover, the interruption of these busy processors may even cause more overhead during the mining process.

\textbf{Summary of merits and limitation.} In this subsection, we have described the key details of some partition-based algorithms for PSPM so far. We further highlight the key differences, advantages and disadvantages between these approaches, as shown in Table \ref{table_disPartition}.

\begin{table*}[!htbp]
	\centering
	\small
	\caption{Apriori-based algorithms for parallel sequential pattern mining.} 
	\label{table_6}
	\newcommand{\tl}[1]{\multicolumn{1}{l}{#1}}   
	\begin{tabular}{|c|l|l|l|l|l|} 
		\hline
		\multicolumn{1}{|c|}{\textbf{Name}} & \multicolumn{1}{|c|}{\textbf{Description}} & \multicolumn{1}{|c|}{\textbf{Pros.}} & \multicolumn{1}{|c|}{\textbf{Cons.}} & \multicolumn{1}{|c|}{\textbf{Year}} \\ \hline

		pSPADE \cite{zaki2001parallel} & 	\multirow{2}{3.5cm}{Parallel SPADE.}  & \multirow{2}{4cm}{$\bullet$ RDLB exploits the dynamic load balancing at both inter-class and intra-class granularities.} & 	\multirow{2}{4cm}{$\bullet$ It mines patterns with level-wise candidate generation-and-test.}  & 2001  \\ 
		& &  & &   \\ 	
		& &  & &   \\ 
		& &  & &   \\ \hline
		
		webSPADE \cite{demiriz2002webspade} & 	\multirow{2}{3.5cm}{A parallel-algorithm variation of SPADE developed for web logs.}  & \multirow{2}{4cm}{$\bullet$ It requires once scan of the input data and multi-processor servers. \\
		$\bullet$  Data and task parallelism are achieved easily by multi-threaded programming. } & 	\multirow{2}{4cm}{$\bullet$ As a Apriori-based, vertical formatting method similar to SPADE, it needs several partial scans of the database.}  & 2002  \\
		& &  & &   \\ 
		& &  & &   \\ 		 
		& &  & &   \\ 
		& &  & &   \\ 
		& &  & &   \\ \hline

		DGSP \cite{yu2015mapreduce} & 	\multirow{2}{3.5cm}{A distributed GSP algorithm based on MapReduce.}  & \multirow{2}{4cm}{$\bullet$ DGSP optimizes the workload balance, partitions the database, and assigns the fragments to Map workers.} & 	\multirow{2}{4cm}{$\bullet$ It may performe poorly because of the repeated scans of the input sequence data.}  & 2005  \\ 
		& &  & &   \\ 
		& &  & &   \\ 
		& &  & &   \\ \hline

		PartSpan \cite{qiao2008partspan} & 	\multirow{2}{3.5cm}{Parallel sequence mining of trajectory patterns.}  & \multirow{2}{4cm}{$\bullet$ It can reduce the I/O cost by using the candidate pruning strategy and reasonable data distribution.} & 	\multirow{2}{4cm}{$\bullet$ It does not scale well and can not efficiently solve the problem of time-consuming and memory cost.}  & 2008  \\ 
		& &  & &   \\ 		 
		& &  & &   \\ 
		& &  & &   \\ \hline

		DPSP \cite{huang2010dpsp}  & 	\multirow{2}{3.5cm}{A sequence algorihtm that mining Distributed Progressive Sequential Patterns on the cloud.}  & \multirow{2}{4cm}{$\bullet$ It performs a progressive mining process in a dynamic database. \\
			$\bullet$ It can discover up-to-date frequent sequential patterns.} & 	\multirow{2}{4cm}{$\bullet$ It may cause the load unbalancing problem because of the required numerous rounds of MapReduce jobs. \\
			$\bullet$ It may easily incur high cost of MapReduce initialization.}  & 2010  \\ 
		& &  & &   \\ 		
		& &  & &   \\ 	
		& &  & &   \\  
		& &  & &   \\ 		
		& &  & &   \\ \hline

		GridGSP \cite{wu2012empirical} & 	\multirow{2}{3.5cm}{A distributed and parallel GSP in a grid computing environment.}  & \multirow{2}{4cm}{$\bullet$ It utilizes the divide-and-conquer strategy. \\
			$\bullet$ Several monitoring mechanisms are developed to help manage the SPM process.} & 	\multirow{2}{4cm}{$\bullet$ Suffer some drawbacks of the GSP algorithm, such as time-consuming and memory cost.}  & 2012  \\ 
		& &  & &   \\ 
		& &  & &   \\ 
		& &  & &   \\ 
		& &  & &   \\ \hline

		SPAMC \cite{chen2013highly} & 	\multirow{2}{3.5cm}{SPAM algorithm on the Cloud.}  & \multirow{2}{4cm}{$\bullet$ Could-based SPAMC utilizes an iterative MapReduce framework to increase scalability.} & 	\multirow{2}{4cm}{$\bullet$ It is not an iterative mining approach, and may incur additional costs for reloading data. \\
			$\bullet$ It does not study the load balancing problem.  \\
			$\bullet$ It is still not scalable enough.}  & 2013  \\ 
		& &  & &   \\ 
		& &  & &   \\ 
		& &  & &   \\ 
		& &  & &   \\ 
		& &  & &   \\ 
		& &  & &   \\ \hline
		
		SPAMC-UDLT \cite{chen2017distributed} & 	\multirow{2}{3.5cm}{SPAM algorithm in the cloud-uniform distributed lexical sequence tree.}  & \multirow{2}{4cm}{$\bullet$ It is an iterative mining approach. \\
			$\bullet$ It studies the load balancing problem. \\
			$\bullet$ Exploiting MapReduce and streaming processes.} & 	\multirow{2}{4cm}{$\bullet$ It does not consider the gap constraint.\\
			$\bullet$ It does not study condense representation (i.e., closed, maximal).}  & 2013  \\ 
		& &  & &   \\ 
		& &  & &   \\ 
		& &  & &   \\ 
		& &  & &   \\ 
		& &  & &   \\ \hline
		
	\end{tabular}
\end{table*}

\subsection{Apriori-based Algorithms for PSPM}

As the above partition-based algorithms of PSPM, input partitioning is not inherently necessary for shared memory systems or MapReduce distributed systems. In this subsection, some typical Apriori-based algorithms for PSPM are introduced. Moreover, the key differences, advantages and disadvantages between these approaches are highlighted at Table \ref{table_6}.

\textbf{pSPADE and webSPADE.} Zaki first extended the efficient SPADE algorithm to the shared memory parallel architecture, called pSPADE \cite{zaki2001parallel}. In the pSPADE framework, input sequence data is assumed to be resided on shared hard drive space, and stored in \textit{lattice} (a vertical data structure). pSPADE utilizes an optimized task parallel formulation approach (Recursive Dynamic Load Balancing, RDLB), and uses two data parallel formulations (single vs. level-wise id-list parallelism and join parallelism), to partition the input spaces. Processes are either assigned id-lists for a subset of sequences, or portions of all id-lists associated with a range of sequences in database. After that, processes collaborate to expand each node in the itemset \textit{lattice}. However, these formulations lead to poor performance due to high synchronization and memory overheads. Zaki then proposed two task distribution schemes, which are able to avoid read/write conflicts through independent search space sub-lattice assignments. Besides, a web-based parallel approach called webSPADE is proposed to analyze the web log \cite{demiriz2002webspade}. webSPADE is also a SPADE-based algorithm to be run on shared memory parallel computers.

\textbf{DGSP and DPSP}. A distributed GSP algorithm based on MapReduce \cite{dean2010mapreduce} framework named DGSP was introduced in \cite{yu2015mapreduce}. The ``two-jobs" structure used in DGSP can effectively reduce the communication overhead for parallelly mining sequential patterns. DGSP optimizes the workload balance, partitions the database, and assigns the fragments to Map workers. All in all, DGSP is a straight-forward extension of GSP in distributed environment, it performs poorly because of the repeated scans of the input data. Huang et al. developed Map/Reduce jobs in DPSP to delete obsolete itemsets, to update current candidate sequential patterns, and to report up-to-date frequent sequential patterns within each period of interest (POI) \cite{huang2010dpsp}. One major drawback of DPSP is that it needs to proceed through numerous rounds of MapReduce's jobs, which will easily cause the load unbalancing problem and incur high cost of MapReduce initialization. Anthor drawback is that it focuses on the POI concept, where the database is divided into many POI windows. Thus, its ability of handling longer sequential patterns is not demonstrated.

\textbf{PartSpan and GridGSP.} After that, some distributed and parallel mining methods, such as parallel sequence mining of trajectory pattern (PartSpan) \cite{qiao2008partspan} and GridGSP \cite{wu2012empirical}, were proposed to find trajectory patterns by extending the traditional GSP algorithm \cite{srikant1996mining}. The PartSpan, which is based on GSP and HPSPM, adopts the prefix-projection and the parallel formulation \cite{qiao2008partspan}. Thus, it can efficiently solve the I/O cost by using the candidate pruning strategy and reasonable data distribution \cite{qiao2008partspan}. GridGSP is a distributed and parallel GSP algorithm in a grid computing environment, it provides a flexible and efficient platform for SPM \cite{wu2012empirical}. In GridGSP, the input sequence data is divided into a number of progressive windows, and then these data independently perform candidate generation on multiple machines. Then, reducer uses the support assembling jobs to count the relative occurrence frequencies of candidate sequences. Besides, GridGSP uses the Count Distribution (CD) technology for counting 1-length candidate sequences, followed by a projection-based task partitioning for solving the remainder of the PSPM problem. The similar strategies were used in PLUTE \cite{qiao2010parallel} and MG-FSM \cite{miliaraki2013mind}. Both PartSpan and GridGSP focus on efficiently mining the frequent patterns under different domain-specific constraints or on different types of data by using MapReduce.

\textbf{SPAMC \& SPAMC-UDLT}. Recently, Huang et al. extend the classic SPAM algorithm to a MapReduce version, named as SPAM algorithm on the Cloud (SPAMC for short). Cloud-based SPAMC utilizes an iterative MapReduce to quickly generate and prune candidate sequences while constructing the lexical sequence tree \cite{chen2013highly}. The core technology of SPAMC is that it mines a sequential database using parallel processing, and the all sub-tasks can be simultaneously distributed and executed on many machines. However, SPAMC does not address the iterative mining problem, as well as the load balancing problem. Thus, reloading data in SPAMC incurs additional costs. Although SPAMC utilizes a global bitmap, it is not scalable enough for handing extremely large-scale databases. To remedy these problems, Chen et al. further devised SPAMC-UDLT to discover the sequential patterns in the cloud-uniform distributed lexical sequence tree, exploiting MapReduce and streaming processes \cite{chen2017distributed}. SPAMC-UDLT dramatically improves overall performance without launching multiple MapReduce rounds, and provides perfect load balancing across machines in the cloud \cite{chen2017distributed}. The results \cite{chen2017distributed} showed that SPAMC-UDLT can remarkably reduce execution time compared to SPAMC, achieve high scalability, and provide much better load balancing than existing algorithms in the cloud.

\textbf{Summary of merits and limitation.}  All the above-described algorithms, which are shown in Table \ref{table_6}, are the Apriori-based algorithms for parallel sequential pattern mining. In general, all these algorithms may also suffer from some drawbacks of serial Apriori-based SPM algorithm. Thus, they do not scale well and can not efficiently solve the problem such as  time-consuming and memory cost.

\begin{table*}[!htbp]
	\centering
	\small
	\caption{Pattern growth algorithms for parallel sequential pattern mining.} 
	\label{table_7}
	\newcommand{\tl}[1]{\multicolumn{1}{l}{#1}}    
	\begin{tabular}{|c|l|l|l|l|l|} 
		\hline
		\multicolumn{1}{|c|}{\textbf{Name}} & \multicolumn{1}{|c|}{\textbf{Description}} & \multicolumn{1}{|c|}{\textbf{Pros.}} & \multicolumn{1}{|c|}{\textbf{Cons.}} & \multicolumn{1}{|c|}{\textbf{Year}} \\ \hline
		
		Par-ASP \cite{cong2005sampling} & 	\multirow{2}{3.5cm}{Parallel PrefixSpan to mine all sequential patterns (ASP).}  & \multirow{2}{4cm}{$\bullet$ It uses a sampling method to accomplish static task partition.} & 	\multirow{2}{4cm}{$\bullet$ The serial sampling component limits the speedup potential when the number of processes increases.}  & 2005  \\ 
		& &  & &   \\ 
		& &  & &   \\ 		
		& &  & &   \\ \hline
		
		Par-CSP \cite{cong2005parallel} & 	\multirow{2}{3.5cm}{Parallel mining Closed Sequential Patterns with selective sampling.}  & \multirow{2}{4cm}{$\bullet$ The first parallel algorithm for mining closed sequential patterns. \\
			$\bullet$ It can reduce processor idle time using a dynamic scheduling scheme. \\
			$\bullet$ Using a load-balancing scheme.} & 	\multirow{2}{4cm}{$\bullet$ The serial sampling component limits the speedup potential when the number of processes increases.}  & 2005  \\ 
		& &  & &   \\ 
		& &  & &   \\ 
		& &  & &   \\ 
		& &  & &   \\ 
		& &  & &   \\ 		
		& &  & &   \\ 
		& &  & &   \\ \hline

		PLUTE \cite{qiao2010parallel} & \multirow{2}{3.5cm}{Parallel sequential pattern mining.}  & \multirow{2}{4cm}{$\bullet$ It focuses on massive	trajectory data, and outperforms PartSpan in mining massive trajectory data.} & 	\multirow{2}{4cm}{$\bullet$ The used prefix projection technology could consume huge memory and a lot of scan time.}  & 2010  \\ 
		& &  & &   \\ 
		& &  & &   \\ 
		& &  & &   \\ 
		& &  & &   \\ \hline
		
		BIDE-MR \cite{yu2012bide}  & \multirow{2}{3.5cm}{A BIDE-based parallel algorithm on MapReduce.} & \multirow{2}{4cm}{$\bullet$ The tasks of closure checking and pruning can be iteratively assigned to different nodes in cluster. \\ $\bullet$ Performed on Hadoop cluster with high speed-ups. \\
			$\bullet$ Can mine closed sequential patterns.} & 	\multirow{2}{4cm}{$\bullet$ Even mining frequent closed sequences does not fully resolve the problem of pattern explosion.}  & 2012  \\ 
		& &  & &   \\ 
		& &  & &   \\ 
		& &  & &   \\ 
		& &  & &   \\ 
		& &  & &   \\ 
		& &  & &   \\ 
		& &  & &   \\ \hline

		DFSP \cite{liao2014dfsp} & 	\multirow{2}{3.5cm}{A Depth-First SPelling algorithm for mining sequences from biological data.} & \multirow{2}{4cm}{$\bullet$ It proposes a three-dimensional list for mining DNA sequences. \\
			$\bullet$ It addresses biological data mining, such as protein DNA analysis.} & 	\multirow{2}{4cm}{$\bullet$ It focuses on PSPM under different domain-specific constraints, but not the general cases without any constraints.}  & 2013  \\ 
		& &  & &   \\ 
		& &  & &   \\ 
		& &  & &   \\ 
		& &  & &   \\ 
		& &  & &   \\ \hline		
		
		\multirow{2}{2cm}{Sequence \\
			-Growth \cite{liang2015sequence}} & 	\multirow{2}{3.5cm}{A MapReduce-based FIM algorithm with sequences.}  & \multirow{2}{4cm}{$\bullet$ It uses a lexicographical order to generate candidate sequences that avoiding expensive scanning processes. \\
			$\bullet$ It provides an extension for trajectory pattern mining.} & 	\multirow{2}{4cm}{$\bullet$ It does not consider the gap-constraint and support hierarchies.}  & 2015  \\ 
		& &  & &   \\ 
		& &  & &   \\ 
		& &  & &   \\ 
		& &  & &   \\ 
		& &  & &   \\ \hline
	\end{tabular}
\end{table*}

\subsection{Pattern Growth Algorithms for PSPM}

As mentioned before, many pattern growth algorithms for serial sequential pattern mining have been extensively studied. Based on these theories and techniques, some of these algorithms have been extended to realize the parallelization. As shown in Table \ref{table_7}. It contains key characteristics, advantages and disadvantages of the pattern growth approaches. The PrefixSpan algorithm also has some parallelized versions, such as Par-ASP \cite{cong2005sampling} and Sequence-Growth \cite{liang2015sequence}. 

\textbf{Par-FP, Par-ASP and Par-CSP}. In order to balance mining tasks, Cong et al. designed several pattern-growth models, Par-FP (Parallel FP-growth) \cite{cong2005sampling}, Par-ASP (Parallel PrefixSpan to mine all sequential-patterns) \cite{cong2005sampling} and Par-CSP (Parallel mining of Closed Sequential Patterns) \cite{cong2005parallel}, to accomplish the static task. Especially, they use a sampling technique, named \textit{selective sampling}, that requires the entire input data be available at each process. Par-ASP is extended by the classical PrefixSpan algorithm \cite{han2001prefixspan}, and the used \textit{selective sampling} in Par-ASP is performed by four steps: sample tree construction, sample tree mining, task partition and task scheduling. After collecting 1-length frequent sequence counts, Par-ASP separates a sample of $k$-length frequent prefixes of sequences in database. It then uses one process to handle the sample via a pattern growth algorithm, recording the execution times for the found frequent sub-sequences. Correspondingly, mining each of frequent sub-sequences can be transformed into a parallel task. A series of projected sub-databases for these frequent sub-sequences are done while estimating their execution times. After that, task distribution is done in the same way as in the Par-FP algorithm \cite{cong2005sampling}. However, it has been found that the serial sampling operations of these algorithms limit the speedup potential when the number of processes increases.

As described at above subsection, some closed-based SPM algorithms have been extensively studied. Unfortunately, no parallel method for closed sequential pattern mining has yet been proposed. In 2005, Cong et al. first proposed the Par-CSP (Parallel Closed Sequential Pattern mining) algorithm \cite{cong2005parallel}  based on the BIDE algorithm. Par-CSP \cite{cong2005parallel} is the first parallel algorithm which aims at mining closed sequential patterns. The process in Par-CSP is divided into independent tasks, this can minimize inter-processor communications. Par-CSP utilizes a dynamic scheduling strategy to reduce processor idle time and is performed in a distributed memory system. It is different from pSPADE using a shared-memory system. Par-CSP also uses a load-balancing scheme to speed up the process. Applying such a dynamic load balancing scheme in a distributed memory system is too expensive to be practical even for big data. In contrast to mine closed sequences, a parallel version of the MSPX \cite{luo2005efficient} algorithm named PMSPX was also proposed. It mines maximal frequent sequential patterns by using multiple samples \cite{luo2012parallel}.

In addition, sequential mining can be applied into biological sequence mining, such as a protein DNA analysis. Recently, the 2PDF-Index \cite{wang2004scalable}, 2PDF-Compression \cite{wang2004scalable}, and DFSP \cite{liao2014dfsp} algorithms were proposed and applied to scalable mining sequential patterns for biological sequences. DFSP \cite{liao2014dfsp} uses a three-dimensional list for mining DNA sequences. It adopts direct access strategy and binary search to index the list structure for enumerating candidate patterns. However, these three works focus on efficiently mining the frequent patterns under different domain-specific constraints or on different types of data (e.g., medical database, DNA sequences), while SPAMC-UDLT \cite{chen2017distributed} focuses on the general cases without any constraints.

\textbf{Summary of merits and limitation.} According the above analytics, it can be concluded that the pattern-growth approaches usually perform better than the early technologies, including the partition-based and Apriori-based algorithms for PSPM. Moreover, the key differences, advantages and disadvantages between these approaches are highlighted in Table \ref{table_6}.

\begin{table*}[!htbp]
	\centering
	\small
	\caption{Hybrid algorithms for parallel sequential pattern mining.}
	\label{table_HybridPSPM} 
	\newcommand{\tl}[1]{\multicolumn{1}{l}{#1}}    
	\begin{tabular}{|c|l|l|l|l|l|} 
		\hline
		\multicolumn{1}{|c|}{\textbf{Name}} & \multicolumn{1}{|c|}{\textbf{Description}} & \multicolumn{1}{|c|}{\textbf{Pros.}} & \multicolumn{1}{|c|}{\textbf{Cons.}} & \multicolumn{1}{|c|}{\textbf{Year}} \\ \hline

		MG-FSM \cite{miliaraki2013mind} & 	\multirow{2}{3cm}{A large-scale algorithm for Frequent Sequence Mining on MapReduce with gap constraints.}  & \multirow{2}{5.5cm}{$\bullet$ Uses some optimization techniques to minimize partition size, computational cost, and communication cost. \\
			$\bullet$ More efficient and scalable than other alternative approaches. \\
			$\bullet$  Any existing SPM method can be used to mine one of its partitions. \\
			$\bullet$ Can handle temporal gaps, maximal and closed sequences, which improves its usability.} & 	\multirow{2}{3.5cm}{$\bullet$ MG-FSM does not support hierarchies, that some of items have hierarchies and frequency can be different.}  & 2013  \\ 
		& &  & &   \\ 
		& &  & &   \\ 
		& &  & &   \\ 
		& &  & &   \\ 
		& &  & &   \\ 
		& &  & &   \\ 
		& &  & &   \\ 
		& &  & &   \\ 		
		& &  & &   \\ \hline

		DFSP \cite{liao2014dfsp} & 	\multirow{2}{3cm}{A Depth-First SPelling algorithm for mining sequences from biological data.} & \multirow{2}{5.5cm}{$\bullet$ It proposes a three-dimensional list for mining DNA sequences. \\
			$\bullet$ It addresses biological data mining, such as protein DNA analysis.} & 	\multirow{2}{3.5cm}{$\bullet$ It focuses on PSPM under different domain-specific constraints, but not the general cases without any constraints.}  & 2013  \\ 
		& &  & &   \\ 
		& &  & &   \\ 
		& &  & &   \\ 
		& &  & &   \\ \hline
		
		\multirow{2}{2cm}{IMRSPM \cite{ge2015mining}}  & 	\multirow{2}{3cm}{Uncertain SPM algorithm in iterative MapReduce framework.} & \multirow{2}{5.5cm}{$\bullet$ An iterative MapReduce framework for mining uncertain sequential patterns.  \\
			$\bullet$ The first work to solve the large-scale uncertain SPM problem.} & 	\multirow{2}{3.5cm}{$\bullet$ It extends the Apriori-like SPM framework to MapReduce, thus suffers from some drawbacks of Apriori-like SPM algorithms.}  & 2015  \\
		& &  & &   \\  
		& &  & &   \\ 
		& &  & &   \\   
		& &  & &   \\    
		& &  & &   \\ \hline

		MG-FSM+ \cite{beedkar2015closing} & 	\multirow{2}{3cm}{An enhanced version of MG-FSM.}  & \multirow{2}{5.5cm}{$\bullet$ A more scalable distributed (i.e., shared-nothing) sequential pattern mining algorithm. \\
			$\bullet$ A more suitable SPM approach to directly integrate the maximality constraint.} & 	\multirow{2}{3.5cm}{$\bullet$ Similar to MG-FSM, MG-FSM+ does not support hierarchies either.}  & 2015  \\ 
		& &  & &   \\ 
		& &  & &   \\ 
		& &  & &   \\ 
		& &  & &   \\ \hline

		LASH \cite{beedkar2015lash} & 	\multirow{2}{3cm}{A LArge-scale Sequence mining algorithm with Hierarchies.}  & \multirow{2}{5.5cm}{$\bullet$ It is the first parallel algorithm for mining frequent sequences with consideration of hierarchies. \\
			$\bullet$ It proposes the pivot sequence miner (PSM) method for mining each partition. \\
			$\bullet$ LASH can search better than MG-FSM because of PSM. \\
			$\bullet$ LASH does not need a global post-processing.} & 	\multirow{2}{3.5cm}{$\bullet$ It cannot handle stream data and uncertain data.}  & 2015  \\ 
		& &  & &   \\ 
		& &  & &   \\ 
		& &  & &   \\ 
		& &  & &   \\ 
		& &  & &   \\ 
		& &  & &   \\ 
		& &  & &   \\ 
		& &  & &   \\ \hline

		\multirow{2}{2cm}{Distributed SPM \cite{ge2016distributed}} & \multirow{2}{3cm}{A distributed SPM approach with dynamic programming.} &  \multirow{2}{5.5cm}{$\bullet$ A memory-efficient approach uses distribute dynamic programming schema. \\
			$\bullet$ Uses an extended prefix-tree to save intermediate results. \\
			$\bullet$ Its time cost is linear.} & 	\multirow{2}{3.5cm}{$\bullet$ It does not consider the gap-constraint and support hierarchies.}  & 2016  \\ 
		& &  & &   \\  
		& &  & &   \\ 
		& &  & &   \\  
		& &  & &   \\ \hline

		\multirow{2}{2cm}{Interesting Sequence Miner (ISM) \cite{fowkes2016subsequence}} & 	\multirow{2}{3cm}{A novel subsequence interleaving parallel model based on a probabilistic model of the sequence database.}  & \multirow{2}{5.5cm}{$\bullet$ It takes a probabilistic machine learning approach (an encoding scheme) to the SPM problem. \\ 
			$\bullet$ All operations on the sequence database are trivially parallelizable.} & 	\multirow{2}{3.5cm}{$\bullet$ It cannot handle stream data and uncertain data.}  & 2015  \\ 
		& &  & &   \\ 
		& &  & &   \\ 
		& &  & &   \\ 
		& &  & &   \\ \hline
		
	\end{tabular}
\end{table*}

\subsection{Hybrid Algorithms for PSPM}

Some hybrid parallel algorithms for SPM are developed by incorporating the above different technologies, as shown in Table \ref{table_HybridPSPM}. Details of some important hybrid algorithms are described below.

\textbf{MG-FSM and MG-FSM+.} MG-FSM \cite{miliaraki2013mind} and its enhanced version MG-FSM+ \cite{beedkar2015closing} are the scalable distributed (i.e., shared-nothing) algorithms for gap-constrained SPM on MapReduce \cite{dean2010mapreduce}. In MG-FSM and MG-FSM+, the notion of w-equivalency w.r.t. ``projected database'' which is used by many SPM algorithms is introduced. Moreover, some optimization techniques are developed to minimize partition size, computational costs, and communication costs. Therefore, MG-FSM and MG-FSM+ are more efficient and scalable than previous alternative approaches. Especially, any existing serial FSM method can be applied into MG-FSM to handle each of its partitions. MG-FSM is divided into three key phases: i) a preprocessing phase, ii) a partitioning phase, and iii) a mining phase; all these phases are fully parallelized. It is more adaptable and flexible that both MG-FSM \cite{miliaraki2013mind} and MG-FSM+ \cite{beedkar2015closing} can handle temporal gaps and maximal and closed sequences, which improves their usability.

\textbf{MG-FSM vs. LASH}. In some real-world applications, hierarchy and frequency of items can be different, while MG-FSM does not support hierarchies. To overcome this drawback, LASH is recently designed for large-scale sequence datasets. It is the first parallel algorithm to discovery frequent sequences by considering hierarchies \cite{beedkar2015lash}. Inspired by MG-FSM, LASH first partitions the input data, and then handles each partition independently and in parallel. Both of them are the distributed algorithms for mining SPs with Maximum gap and maximum length constraints using MapReduce \cite{dean2010mapreduce}, they are all more adaptable and flexible than other PSPM algorithms. Notice that LASH does not have a global post-processing, it divides each sequence by pivot item and performs local mining (PSM). Therefore, LASH can have better search performance than MG-FSM and MG-FSM+. 

\textbf{ACME}. For the application in bioinformatics, Sahli et al. proposed ACME \cite{sahli2013parallel}, which applies tree structure to extract longer motif sequences on supercomputers. ACME is a parallel combinatorial method for extracting motifs repeated in a single long sequence (e.g., DNA) instead of multiple long sequences. It proposes two novel models, CAST (cache aware search space traversal model) and FAST (fine-grained adaptive sub-tasks), to effectively utilize memory caches and processing power of multi-core shared-memory machines. ACME introduces an automatic tuning mechanism that suggests the appropriate number of CPUs to utilize. In particular, it is large-scale shared nothing systems with tens of thousands of processors, which are typical in cloud computing. It has been shown that ACME achieves improvement in serial execution time by almost an order of magnitude, especially supports for longer sequences (e.g., DNA for the entire human genome).

\textbf{Summary of merits and limitation.} In most PSPM algorithms, they are hybrid inherently. Although the hybrid methods for PSPM incorporate different technologies to effectively reduce communication cost, memory usage and execution time, which usually have a significant performance improvement than the above partition-based, Apriori-based, and pattern growth algorithms. There is no doubt that they also have their advantages and disadvantages. Details of characteristics, advantages and disadvantages about these hybrid methods for PSPM are described at Table \ref{table_HybridPSPM}.

\section{Advanced Topics in PSPM} 

In this section, some advanced topics of parallel sequential pattern mining are provided and discussed in details, including parallel quantitative / weighted / utility sequential pattern mining, parallel sequential pattern mining from uncertain data and stream data, hardware accelerator for PSPM. For specific problems and tasks in real-world applications, these advanced algorithms can provide more choices and options to user.

When someone is beginning his/her way in data mining, especially in sequential pattern mining, he/she often face the problem of choosing the most appropriate algorithm for his/her specific task. There is no doubt that no one solution or one approach that fits all. In general, some problems may be very specific and require a unique approach. In the above sections, we have provided a detailed survey and discussion for the different categories of PSPM so far. In our opinions, people who want to use the PSPM algorithm should first understand their data, than categorize the problem or task, and understand the specific constraints and the requirements/goals (e.g., business requirement, accuracy, efficiency, maximal memory usage, and scalability). Finally, they can find the available algorithms for his/her specific requirement.

\subsection{Parallel Quantitative / Weighted / Utility Sequential Pattern Mining}

Generally speaking, most PSPM algorithms focus on how to improve their efficiency. In many cases, effectiveness can be far more important than efficiency. It is commonly seen that sequence data contains various quantity \cite{hong1999mining}, weight \cite{lan2014efficient,yun2008new,lin2015rwfim}, or utility with different items/objects \cite{lin2015efficient,gan2018survey,2gan2018survey}. For example, consider some transactions from shopping baskets. As shown in Table \ref{table_seq1}, the quantity and price of each item are not considered in this database. In general, some useful information (i.e., quantitative, weight and utility) are simply ignored in the current form of sequential pattern mining. Up to now, some approaches have been proposed to handle quantity, weight or utility constraints in sequential pattern mining, and a literature review has been provided by Gan at al. in \cite{2gan2018survey}. Many sequential pattern algorithms have been introduced, such as quantitative sequential pattern \cite{kim2007squire,hong1999mining}, weighted sequential pattern \cite{lan2014efficient,yun2008new}, utility-driven sequential pattern \cite{alkan2015crom,2gan2018survey}, etc. All of these efforts rely on sequence data and are not able to scale to larger-scale datasets. How to achieve parallelism of these methods is an open issue and challenge. 

Recently, the methods for mining above useful sequential patterns are extended to achieve parallelism, such as the PESMiner (Parallel Event Space Miner) framework \cite{ruan2014parallel} for parallel and quantitative mining of sequential patterns at scale, and the Spark-based BigHUSP algorithm \cite{zihayat2016distributed} for mining high-utility sequential patterns. As one of the big data models for utility mining \cite{2gan2018survey}, BigHUSP is designed based on the Apache Spark platform \cite{zaharia2012resilient} and takes advantage of several merit properties of Spark such as distributed in memory processing, fault recovery and high scalability \cite{zihayat2016distributed}. On one hand, both PESMiner \cite{ruan2014parallel} and BigHUSP \cite{zihayat2016distributed} can identify informative sequential patterns with respect to a business objective, such as interval-based quantitative patterns and profitable patterns. Thus, they may be more effective than those which simply ignore the useful information. On the other hand, these frameworks integrate domain knowledge, computational power, and MapReduce \cite{dean2010mapreduce} or the Apache Spark \cite{zaharia2012resilient} techniques together to achieve better feasibility, quality and scalability.

\subsection{PSPM in Uncertain Data}
In many real-life applications, uncertain data is commonly seen, and uncertain sequence data is widely used to model inaccurate or imprecise timestamped data \cite{muzammal2015mining}, while traditional datat mining (i.e., FIM, ARM, SPM) algorithms are inapplicable when handling uncertain data. Other related algorithms for parallel sequential pattern mining are still being developed, such as the PSPM of large-scale databases by using a probabilistic model \cite{fowkes2016subsequence}. Recently, Ge at al. proposed an iterative MapReduce framework for mining uncertain sequential patterns, where the new iterative MapReduce-based Apriori-like uncertain SPM framework is quite different from the traditional SPM algorithms \cite{ge2015mining}. An uncertain sequence database $D$ = $ \{d_1, ..., d_n\} $ is abstracted by a Resilient Distributed Datasets (RDD) \cite{zaharia2012resilient} in Spark. Uncertain sequences in the RDD are allocated to a cluster of machines and can be processed in parallel. A sequential pattern $s$ in $D$ is called a probabilistic frequent pattern (\textit{p-FSP}) if and only if its probability of being frequent is at least $\psi_p$, denoted by $ P(sup(s) \geq \psi_s) \geq \psi_p$. Here, $\psi_s$ is the user-defined minimum probability threshold. However, the frequentness of $s$ in an uncertain database is probabilistic. 

For application of SPM in uncertain databases \cite{muzammal2015mining}, it is challenging in terms of efficiency and scalability. Ge and Xia further developed a Distributed Sequential Pattern (DSP) mining algorithm in large-scale uncertain databases based on Spark, which relies on a memory-efficient distributed dynamic programming (DP) approach \cite{ge2016distributed}. Although MapReduce \cite{dean2010mapreduce} is widely used for processing big data in parallel, while it does not support the iterative computing model, its basic framework cannot be directly used in SPM. Directly applying the DP method to Spark is impractical because its memory-consuming characteristic may cause heavy JVM garbage collection overhead in Spark \cite{ge2016distributed}. For bioinformatics, a new spark-based framework was proposed to mine sequential patterns from uncertain big deoxyribonucleic acid (DNA) \cite{hassani2011towards}.

\subsection{PSPM in Stream Data} 

Most of the above PSPM algorithms are developed to handle traditional database management system (DBMS)\footnote{\url{https://en.wikipedia.org/wiki/Database}} where data are stored in finite, persistent databases. For some real-life applications, it is commonly seen to process the continuous data stream which is quite different from DBMS. Stream data is temporally ordered, fast changing, massive, and potentially infinite \cite{harries1998extracting}. Few stream algorithms have been developed for parallel mining sequential patterns up till now. Recently, Chen et al. proposed two parallel sequential pattern mining algorithms, SPAMC \cite{chen2013highly} and SPAMC-UDLT \cite{chen2017distributed}, for processing stream data. Since SPAMC is still not scalable enough for mining large-scale data, the SPAMC-UDLT algorithm \cite{chen2017distributed} was further proposed to address the fast changing massive stream data. Notice that the characteristics of SPAMC and SPAMC-UDLT have been highlighted at Table \ref{table_6}, and some comments and differences about them have already been mentioned before. The advantages of these cloud-based algorithms are that they can scan the stream data only once, and can effectively generate all the frequent sequential patterns by using stream mining operation. In cases of multiple streams in parallel, the MSSBE algorithm \cite{hassani2011towards} can find sequential patterns in a multiple-stream environment, where pattern elements can be part of different streams.

\subsection{Hardware Accelerator for PSPM}

Generally, many efforts have been made to speed up SPM via software and hardware. On the one hand, parallel implementation is the general way. On the other hand, hardware accelerators allow a single node to achieve orders of magnitude improvements in performance and energy efficiency. As mentioned before, Graphics processing units (GPUs) \cite{anderson2008general, boggan2007gpus} have attracted much attention due to their cost-effectiveness and enormous power for massive data parallel computing. General-purpose GPUs leverage high parallelism, but GPUs' single instruction multiple data (SIMD) and lockstep organization means that the parallel tasks must generally be similar. Lin et al. developed a novel parallel algorithm on GPUs for accelerating pattern matching \cite{lin2013accelerating}. Based on the traditional AC algorithm, they implement three CPU versions and one GPU version as follows: ACCPU, DPACOMP, PFACOMP, and PFACGPU \cite{lin2013accelerating}.

Then Memeti and Pllana introduced an approach for accelerating large-scale DNA sequence analysis and SIMD parallelism using many-core architectures (such as GPU, Intel Xeon Phi coprocessor) \cite{memeti2015accelerating}. Recently, Wang et al. proposed a hardware-accelerated solution, AP-SPM and GPU-GSP, for sequential pattern mining (SPM) \cite{wang2016sequential}. The major ideas are summarized as follows: It uses Micron's new Automata Processor (AP), which provides native hardware implementation of non-deterministic finite automata. Based on the well-known GSP approach, AP SPM derives a compact automaton design for matching and counting frequent sequences. According to experiments with a single-threaded CPU, multi-core CPU, and GPU GSP implementations \cite{wang2016sequential}, it has been shown that the AP-accelerated solution outperforms PrefixSpan \cite{han2001prefixspan} and SPADE \cite{zaki2001spade} on multi-core CPU, and scales well for larger datasets.

\section{Open-Source Software} 
Although the problem of sequential pattern mining has been studied for more than two decades, and the advanced topic of parallel sequential pattern mining also has been extended to many research fields, few implementations or source code of these algorithms are released. This brings some barriers to other researchers that they need to re-implement algorithms for using the algorithms or comparing the performance with novel proposed algorithms. To make matter worse, it may cause the unfairness while running experimental comparison, since the performance of pattern mining algorithms (i.e., FIM, ARM and SPM) may commonly depending on the used compiler and the used machine architecture. We provide some open-source software of SPM and PSPM in Table \ref{table_software}.

\begin{table*}[!htbp]
	\centering
	\caption{Open-source software for SPM or PSPM.}
	\label{table_software}
	\newcommand{\tl}[1]{\multicolumn{1}{l}{#1}}    
	\begin{tabular}{|c|l|l|l|} 
		\hline
		\multicolumn{1}{|c|}{\textbf{Name}} & \multicolumn{1}{|c|}{\textbf{Contributors}} & \multicolumn{1}{|c|}{\textbf{Website}} \\ \hline
		
		Sequence Mining  &  Zaki et al.  & \url{http://www.cs.rpi.edu/~zaki/www-new/pmwiki.php/Software} \\ \hline	
		
		SPAM \cite{ayres2002sequential} & 	Gehrke et al.  & \url{http://himalaya-tools.sourceforge.net/Spam} \\ \hline
		
		SPMF \cite{srikant1996mining} &  Fournier-Viger et al.  & \url{http://www.philippe-fournier-viger.com/spmf} \\ \hline	
		
		MLlib  &   Apache Spark  & \url{https://spark.apache.org/mllib/} \\ \hline	
		
		MG-FSM \cite{masseglia2000efficient} & 	Miliaraki et al.  & \url{https://github.com/uma-pi1/mgfsm} \\ \hline
		
		LASH \cite{beedkar2015lash}  &  Beedkar et al.  & \url{http://uma-pi1.github.io/lash} \\ \hline		
	\end{tabular}
\end{table*}

\textbf{$\bullet$ Sequence mining}. Zaki releases many implementations or source code of data mining algorithms on his Website, including itemset mining and association rules, sequence mining, itemset and sequence utilities, tree mining, graph mining \& indexing, and clustering. He provides C code for three sequence mining algorithms.

\textbf{$\bullet$ SPAM}. There is a special website (as shown in Table \ref{table_software}), named ``Himalaya Data Mining Tools", built for the classical SPAM algorithm \cite{ayres2002sequential} which aims at finding all frequent sequences. It provides an overview, an illustrative example, source code and documentation, as well as the presentation slides of SPAM.

\textbf{$\bullet$ SPMF}. SPMF \cite{fournier2016spmf} is a well-known open-source data-mining library, which implements many algorithms and has been cited in more than 600 research papers since 2010. SPMF is written in Java, and provides implementations of 120 data mining algorithms, specialized in sequential pattern mining. SPMF has the largest collection of implementations of various algorithms for sequential pattern mining. It provides the Java code for AprioriAll \cite{agrawal1995mining}, GSP \cite{srikant1996mining}, PrefixSpan \cite{han2001prefixspan}, SPADE \cite{zaki2001spade}, SPAM \cite{ayres2002sequential} and many others. The only problem is that SPMF does not provide any parallel algorithms.

\textbf{$\bullet$ MLlib}. Machine Learning Library (MLlib) is a scalable Apache-Spark-based machine learning library. MLlib contains many algorithms and utilities, including frequent itemset mining, association rule mining, sequential pattern mining, as well as many other algorithms for other mining tasks. Algorithms in MLlib are written in Java, Scala, Python, and R. Up to now, MLlib has become an important and widely used framework with SPM algorithms.

\textbf{$\bullet$ MG-FSM \& MG-FSM+}. MG-FSM \cite{miliaraki2013mind} is a scalable, general-purpose (e.g., general, with Maximum gap constraints, and with maximum length constraints, etc.) frequent sequence mining algorithm built for MapReduce. The Java implementation and command line options of MG-FSM are provided at Github.

\textbf{$\bullet$ LASH}. LASH \cite{beedkar2015lash} is a new high scalable, MapReduce-based distributed sequence mining algorithm. It mines sequential patterns by considering items' hierarchies. The Java source code of LASH is provided at Github, including prerequisites for building LASH and running instructions.

\section{Challenges and Opportunities in Parallel Sequential Pattern Mining} 
\label{sec:sparse-input-dt}
\label{sec:4}

\subsection{Challenges in PSPM}
In recent decades, many models and algorithms have been developed in data mining to efficiently discover the desired knowledge from various types of databases. There are still, however, many challenges in big data mining. According to Labrinidis and Jagadish's study, the major challenges in big data analysis include: data inconsistency, data incompleteness, scalability, timeliness, and security \cite{labrinidis2012challenges}. Although useful developments have been made in the field of parallel computing platforms, there are still some technical challenges \cite{chang2008bigtable, tanenbaum2007distributed}. Despite advances in the field of parallel data mining, especially parallel frequent itemset mining and parallel sequential pattern mining, both in academia and industry, significant challenges still remain in parallel sequential pattern mining. Some challenges which need to be dealt with for realizing parallel sequential pattern mining are respectively explored in the following section.

\textbf{(1) \textit{Complex types of sequence data}.} Sequential data commonly occurs in many fields, current developments in parallel sequential pattern mining have successfully addressed some types of sequence data, including the general type of sequence, sequence data containing quantity / weight / utility \cite{2gan2018survey}, uncertain sequential data \cite{ge2015mining}, stream data \cite{chen2017distributed}, and so on. The related algorithms and some extensions have been mentioned previously. However, there are still various more complex types of sequence data in a wide range of applications. For example, the spatio-temporal sequential data contains rich semantic features and information (i.e., time-space information, location, spatial and temporal relationships among data points of trajectories) than the traditional sequential data. How to improve the mining efficiency, and how to extract more rich information rather than the support-based occurrence are more interesting and helpful to real-world applications. For future research, it is a big challenge to develop more adaptable and flexible PSPM frameworks for dealing with more complex types of sequence data in a parallel environment.

\textbf{(2) \textit{Multi-modal data}.} In the big data era, a key property of multi-modality is complementarity. Each modality brings to the whole some type of added value that cannot be deduced or obtained from any of the other modalities in the setup. In mathematical field, this added value is so called \textit{diversity}. How to integrate different knowledge from the \textit{diversity} of multi-modal data using parallel sequential pattern mining algorithms is a potential challenge in big data area.

\textbf{(3) \textit{Dynamic sequence data}.} In a wide range of applications, the processed data may be commonly dynamic but not static \cite{harries1998extracting}. The dynamic data is more complicated and difficult than the static data. Although some approaches have been developed for handling dynamic sequence data, such as incremental mining \cite{cheng2004incspan,el2004fs,zhang2014maintaining}, decremental mining \cite{lin2014efficiently}, online progressive mining \cite{huang2008general}, and stream data processing \cite{harries1998extracting,chen2017distributed}. Most existing studies are designed without parallelization, research on parallelly mining sequences from dynamic sequence data has seldom been done so far. Therefore, there are still some challenges when parallelizing these mining methods, especially for parallelizing dynamic sequential pattern mining algorithms.

\textbf{(4) \textit{Scalability}.} In general, parallelizing SPM algorithms for dealing with big data is more complicated and difficult than that of small data. It easily brings some challenging problems, such as availability, accuracy, and scalability. When dealing with big data, the disadvantages of current data mining techniques are centered on inadequate scalability and parallelism. Scalability is one of the core technologies to meet these challenges.  Although a significant amount of developments, such as MG-FSM, LASH, and MLlib, have been reported, there is still much improvement for parallel implementation. There is still, however, a major challenge to realize the needed scalability of the developed PSPM algorithm with demonstrated elasticity and parallelism capacities.

\textbf{(5) \textit{Privacy}.} Data has the inestimable value (i.e., hidden knowledge and more valuable insights). Data mining and analytics offer many useful tools and play an important role in many fields. However, a major risk in data management and data mining is data leakage, which seriously threatens privacy. Public concerns regarding privacy are rising, which is a major concern in data mining, especially big data mining. On one hand, policies that cover all user privacy concerns should be developed. On the other hand, privacy preserving data mining (PPDM) \cite{zhu2017differentially} and privacy preserving utility mining (PPUM) \cite{gan2018privacy} are also needed. Up to now, many technologies and algorithms have been developed for PPDM \cite{zhu2017differentially} or PPUM \cite{gan2018privacy}, but few of them are related to parallel sequential pattern mining. With the developments of SPM to analyze personal data, it is quite necessary and challenging to address the privacy preserving problem for parallel sequential pattern mining.

\subsection{Opportunities in PSPM}
There are a variety of traditional, distributed, and parallel data mining algorithms in numerous real-life applications. To meet the demand of large-scale and high performance computing, the problem of parallel/distributed data mining has received considerable attention over the past decade. There are many types of parallel and distributed systems, such as multi-core computing \cite{ranger2007evaluating,dolbeau2007hmpp,huynh2017efficient}, grids \cite{liu2005agent, luo2007distributed}, peer-to-peer (P2P) systems \cite{rao2010optimal}, ad-hoc networks \cite{xue2006optimal}, cloud computing systems \cite{gkatzikis2013migrate}, and the MapReduce framework \cite{dean2010mapreduce}. All these developments can provide theoretical and technical support for parallel sequential pattern mining. Some opportunities are summarized below.

\textbf{(1) \textit{New applications}}. According to the research of current status, most algorithms of PSPM focus on how to improve the mining efficiency. It is commonly seen that sequential data occurs in many fields such as basket analysis, sensor networks, bioinformatics, social network analysis, and the Internet of Things. There are some important research opportunities for PSPM in these fields. Instead of focusing on faster general PSPM algorithms, the trend clearly goes towards to extend parallel sequential pattern mining in new applications, or in new ways for existing applications \cite{fournier2017survey}. For example, the utility-driven mining \cite{2gan2018survey} on large-scale sequence data is more practical but challenge.

\textbf{(2) \textit{Advanced parallel computing environment}}. To overcome the scalability challenge of big data, several attempts have been made in exploiting massive parallel processing architectures, e.g., cloud \cite{gkatzikis2013migrate}, MapReduce \cite{dean2010mapreduce}, Spark \cite{zaharia2012resilient}. All these computing infrastructures provide new opportunities for the design of parallel data mining, especially for parallel sequential pattern mining algorithms. To some degree, all these developments can provide the necessary theoretical and technical support for parallel sequential pattern mining.

\textbf{(3) \textit{Developments from hardware and software}}. As mentioned before, a trend of parallel computing is graphics processing unit (GPU) based parallel techniques. Architecturally, a GPU is composed of hundreds of cores that can simultaneously handle thousands of threads/tasks \cite{anderson2008general, boggan2007gpus}. More and more supercomputers have the GPU component which is different from the traditional CPU. Therefore, we can expect an enhancement on parallel data mining with the progress of powerful computational capabilities. For future research, the developments from hardware and software strongly provide some opportunities for the sequence mining task in a parallel environment.

\textbf{(4) \textit{Keeping pattern short and simple}}. Although there have been numerous studies on pattern mining (e.g., FIM, ARM, SPM, etc.), the study on interesting pattern mining is still limited. Most of them are only limited to discover the complete set of desired patterns, a huge pattern mining results may still confuse users to make inefficient decisions. What iss worse, they may easily cause redundant results when mining long patterns or with a low support threshold. Returning a condensed or more constrained set is an interesting issue. Some methods of condensed SPM (i.e., CloSpan \cite{yan2003clospan}, BIDE \cite{wang2004bide}, MSPX \cite{luo2005efficient} and ClaSP \cite{gomariz2013clasp}, as shown at Table \ref{table_4}) or summarization SPM (i.e., summarizing event sequences \cite{tatti2012long}, SQUISH \cite{bhattacharyya2017efficiently}) have been proposed, that is, instead of returning all patterns that occur more often than a certain threshold, these methods only return a condensed or more constrained set. How to achieve parallelism of these condensed or summarizing methods is an open issue and challenge.

\textbf{(5) \textit{Utilizing deep learning technology}}. Deep learning \cite{lecun2015deep} is a hot topic in current research. There are some works focus on deep learning for sequential data (i.e., text, EEG data and DNA), such as Recurrent Neural Networks (RNN) \cite{williams1989learning}, and Long Short Term Memory (LSTM) \cite{hochreiter1997long}. Unlike the standard neural network, RNN allows us to operate over sequences of vectors \cite{williams1989learning}. And RNN uses internal memory to process arbitrary sequences of inputs. LSTM is particularly usable by a category of learning machines, adapted to sequential data \cite{hochreiter1997long}. Many deep learning algorithms have been proposed to model sequential data, and then to achieve different learning goals. Despite their flexibility and power, how to make them parallelized for sequential pattern mining can lead to challenges and open many new research fields.

\textbf{(6) \textit{Other important issues}}. Beyond the above points, there are some other important issues that can be reasonably considered and further developed, such as visualization techniques and the privacy issue for parallel sequential pattern mining in the big data era. How to design efficient and more flexible PSPM approachs to support iterative and interactive mining is also an interesting issue.

\section{Conclusion} \label{sec:conclusion}

Typically, sequential pattern mining algorithms aim at discovering the desired frequent sequential patterns. Since traditional data mining algorithms generally have some problems and challenges while  processing large-scale data, parallel data mining has emerged as an important research topic. However, fewer studies summarized the related development in parallel sequential pattern mining in parallel computing environments, or summarized them as a taxonomy structure. In this paper, we first review the key characteristics about some parallelism methods, and then highlight differences and challenges between parallel computing platforms and distributed systems. We then highlight and discuss some related works on parallel sequential pattern mining in several categories. The main contributions are that we investigate recent advances in parallel sequential pattern mining and provide the status of the field in detail, including sequential pattern mining (SPM), parallel frequent itemset mining (PFIM), and parallel sequential pattern mining (PSMP). Both basic algorithms and advanced algorithms for parallel sequential pattern mining are reviewed in several categories, the key ideas, advantages and disadvantages of each approach are also pointed out in details. We further provide some related open-source software of PSPM, that may reduce barriers from research and algorithm implementation. Finally, we briefly point out some challenges and opportunities of parallel sequential pattern mining for future research.

\section{Acknowledgment}

We would like to thank the anonymous reviewers for their detailed comments and constructive suggestions for this paper. This research was partially supported by the National Natural Science Foundation of China (NSFC) under Grant No.61503092, and by the China Scholarship Council Program.


\bibliographystyle{ACM-Reference-Format-Journals}
\bibliography{paper}

\end{document}